\documentclass[aps, floatfix, a4paper, nofootinbib,
superscriptaddress, 11pt]{revtex4}

\usepackage[english]{babel}
\usepackage[T1]{fontenc}
\usepackage[utf8]{inputenc} 
\usepackage{graphicx,float,tikz}
\usepackage[all]{xy}
\usepackage{amsmath,upgreek}
\usepackage{amsfonts}
\usepackage{amsthm}   
\usepackage{bbold}
\usepackage{bm}
\usepackage{amssymb}
\usepackage{color}
\usepackage{epsfig}		

\usepackage{subfigure}
\usepackage{pdfpages}
\usepackage{dsfont}

\usepackage{multirow}
\usepackage{silence}
\usepackage[colorlinks=true, linkcolor=blue, citecolor=blue, urlcolor=blue]{hyperref}

\definecolor{green1}{RGB}{0,128,0}

\usepackage{bookmark,textgreek}

\begin{document}

\title{ Quantization Rules in Holographic QCD Models}

\author{Luiz F. Ferreira}
\email{lffaulhaber@gmail.com}
\affiliation{Instituto de Física y Astronomía, Universidad de Valparaíso, A. Gran Bretana 1111}
\author{Alfredo Vega}
\email{alfredo.vega@uv.cl}
\affiliation{Instituto de Física y Astronomía, Universidad de Valparaíso, A. Gran Bretana 1111}

\begin{abstract}

In this paper, we investigate quasinormal modes in holographic QCD models from the perspective of the WKB approximation. We derive the generalized Bohr–Sommerfeld quantization rules for quasi-stationary states in holographic QCD models. As a simple application of these formulas, we compute the quasinormal modes of scalar and vector fields in the soft-wall model, where an analytic expression for the real part of the frequency can be found in the low-temperature regime. Additionally, this study provides useful insights into the dissociation process in holographic QCD models.

\end{abstract}


\maketitle

\newpage
\medbreak

\section{Introduction}

Quasinormal modes of black holes have attracted considerable interest in recent years, emerging as an important research topic within gravitational physics. These modes correspond to characteristic oscillations intrinsic to black holes and exhibit dissipative behavior due to the presence of the event horizon. Typically, quasinormal modes are described by complex frequencies, where the real part relates to the oscillation frequency and the imaginary part governs the decay rate of perturbations. The eigenfunctions associated with quasinormal modes are generally non-normalizable and do not constitute a complete basis, highlighting the complex mathematical structure underlying these oscillations. Comprehensive reviews of quasinormal modes can be found, for instance, in \cite{Kokkotas:1999bd,Nollert:1999ji,Berti:2009kk,Konoplya:2011qq}.

Within the framework of gauge/gravity duality, quasinormal modes  play an essential role in investigating both equilibrium and non-equilibrium phenomena in strongly coupled gauge theories at finite temperature. This significance stems from the fact that the gauge theory on the boundary is holographically dual to a higher-dimensional gravitational bulk containing black holes or black branes and their fluctuations. Specifically, the quasinormal modes spectrum of a given gravitational background corresponds exactly to the pole positions (in momentum space) of the gauge theory’s retarded correlators. Thus, quasinormal modes provide valuable insights into the quasiparticle spectra and transport coefficients of the gauge theory. As a result, the analysis of quasinormal modes has become a standard and powerful technique for probing the near-equilibrium dynamics of gauge-theory plasmas with holographic duals.

\begin{figure}[!htb]
	\centering
	\includegraphics[scale=0.55]{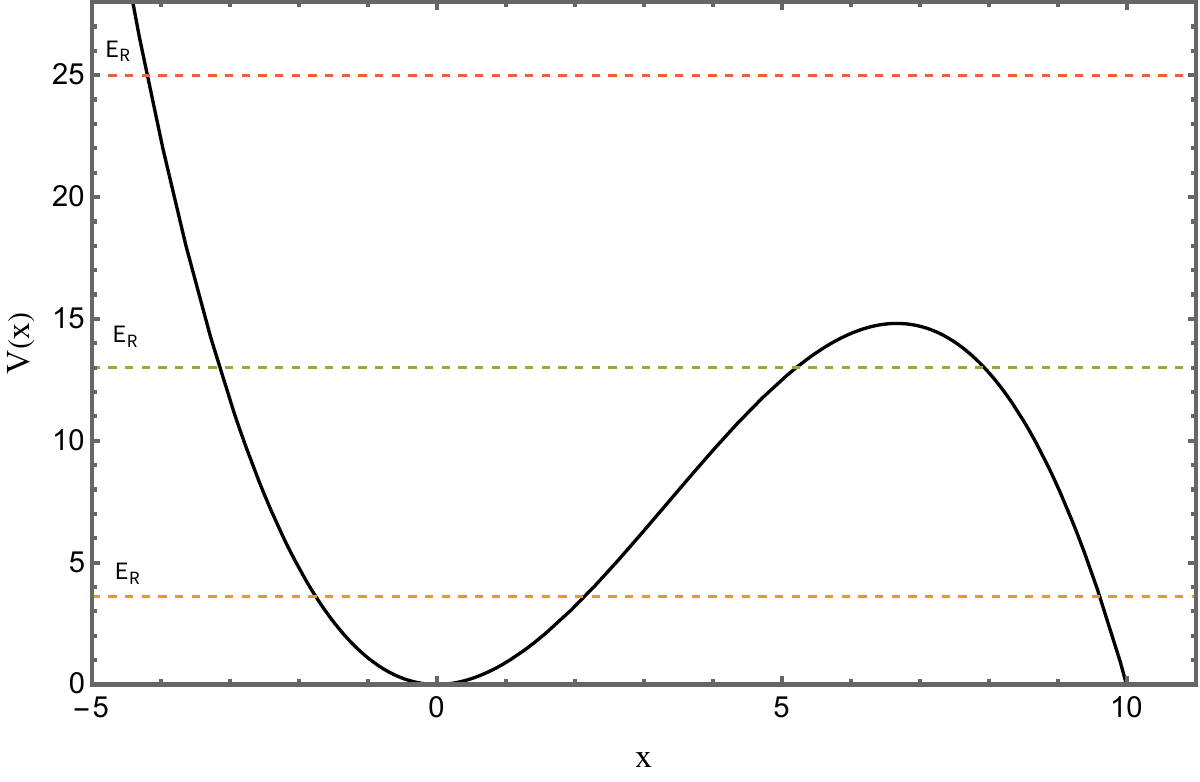}
	\caption{  A quantum mechanical potential with a barrier that exhibits quasi-stationary  states. The dashed lines indicate the region where these states may be found. }
    \label{Barrier Potential}
\end{figure}

In the conformal AdS/CFT case, quasinormal mode frequencies are known to scale linearly with temperature~\cite{Son:2002sd,Nunez:2003eq,Kovtun:2005ev}. By contrast, holographic QCD models~\cite{Karch:2006pv,Colangelo:2007pt,Colangelo:2008us,Gubser:2008ny,Gubser:2008sz,Gursoy:2008bu,Gursoy:2008za,Gursoy:2009jd,Gursoy:2010fj,Branz:2010ub,Li:2011hp,Kajantie:2011nx,Gutsche:2011vb,Alho:2012mh,Cai:2012xh,Li:2013oda,Finazzo:2014zga,Bohra:2020qom,Ferreira:2020iry,Gutsche:2019blp,Gutsche:2019pls,Arefeva:2022avn,Mamani:2022qnf,Arefeva:2024xmg}, which incorporate a mass gap~\cite{Gursoy:2007er}, exhibit deviations from this behavior. At zero temperature, the equations governing excitations can be reduced to a Schrödinger-like form with a confining potential and an infinite barrier at the boundary, resulting in a discrete hadronic spectrum. At non-zero temperature, a finite barrier forms, as illustrated in Fig.~\ref{Barrier Potential}. In this regime, tunneling through the barrier becomes possible, giving rise to metastable, localized states with finite lifetimes. Consequently, the real part of the quasinormal mode frequency represents the mass of the associated quasi-bound state, converging to the corresponding hadronic vacuum mass as the temperature approaches zero. Meanwhile, the imaginary part of the frequency characterizes the width of the quasiparticle, determining its lifetime and capturing the dissipative dynamics intrinsic to strongly coupled gauge theories.

The similarity to quantum mechanics suggests employing the WKB  approximation to study the spectrum of quasinormal modes. In the context of atomic physics, a generalized Bohr–Sommerfeld quantization rule incorporating the effects of a potential barrier (as illustrated in Fig. \ref{Barrier Potential}) was developed using the WKB approximation in \cite{Popov1968QuasiclassicalAF,Popov1991QuantizationRW,Popov1991QUANTIZATIONRF,Mur1993QuantizationRF,Mur1993TheWM}, extending the classical condition originally formulated for bound states. Inspired by this similarity, we derive a generalized Bohr–Sommerfeld rule for quasi-stationary states in holographic QCD models. As in quantum mechanics, quasinormal modes can be categorized into three regimes based on their localization: near the potential minimum, in the vicinity of the top of the potential barrier, and in the above-barrier region.

This paper is organized as follows. In Sec. II, we revisit the effective holographic potential in the soft-wall model. In Sec. III, we derive the generalized BS  quantization rule and apply it to compute quasinormal modes  of a scalar field in the soft-wall background. Sec. IV is devoted to the computation of quasinormal modes for a vector field in the same model. Finally, our conclusions and outlook are presented in Sec. V.

\section{The Effective Holographic Potential}

In this section, we revisit the derivation of the effective potential within the framework of holographic QCD models. For simplicity, we focus on the soft-wall model and examine the glueball case at rest, which is related to the scalar field on the gravity side. The procedure outlined here can be readily extended to other fields in holographic QCD models.

\subsection{ Scalar field in the soft-wall model}

The Schwarzschild black hole in $AdS_5$ is described in the Poincaré patch by the following  metric:
\begin{equation}\label{metricT}
    ds^2=\frac{R^2}{z^2}\left( -f(z)dt^2+d\vec{x}\cdot  d\vec{x} +\frac{dz^2}{f(z)}    \right)\,,
\end{equation}
where $f(z)=1-\frac{z^4}{z_{h}^{4}}$, $z_h$ denotes the position of the horizon, and $R$ represents the curvature radius of $AdS_5$. The black hole temperature is given by:
\begin{equation}\label{Te}
    T=\frac{1}{4\pi}\left|f'(z_h) \right|= \frac{1}{\pi z_h}\,.
\end{equation}
The bulk action for the massless scalar field $\phi$, which is assumed to be dual to the gauge theory operator $Tr (F^2)$, is given by
\begin{equation}\label{acs}
    S=-\int d^5x \sqrt{-g}e^{-\Phi}g^{MN}\partial_M \phi \partial_N \phi
\end{equation}
where $\Phi=c^2z^2$ represents the dilaton field. The equation of motion derived from the action \eqref{acs} and the metric \eqref{metricT} is
\begin{equation}\label{eqz0}
    \partial_{\mu}\left(\sqrt{-g}e^{-\Phi}g^{\mu\nu}\partial_{\mu} \phi \right) = 0.
\end{equation}
Assuming a plane-wave ansatz solution of the form $ \phi (\vec{x},z)= e^{-i\omega t}\phi (z)$, the equation of motion (\ref{eqz0}) reduces to:
\begin{equation}\label{eqz1}
    \phi''+\phi'\left( \frac{f'}{f} - \frac{3}{z}-\Phi'\right)+\omega^{2} \phi = 0.
\end{equation}
where the prime (') denotes the derivative with respect to $z$. To simplify the calculations, we set $c=1$. Thus, the equation of motion (\ref{eqz1}) transforms into:
\begin{equation}\label{eqz2}
    \phi''+\phi'\left( \frac{f'}{f} - \frac{3}{z}-2z\right)+\omega^{2} \phi = 0.
\end{equation}

In order to study the holographic potential in the soft wall model, we need to transform Eq. (\ref{eqz1}) into a Schrödinger-like equation. By applying the Liouville transformation:
\begin{eqnarray} \label{TL}
\psi &=& e^{-B/2} \, \phi \,, \\[1ex]
r_* &=& \frac{1}{2} z_h \left[ - \arctan\left( \frac{z}{z_h} \right) + \frac{1}{2} \ln\left( \frac{z_h - z}{z_h + z} \right) \right] \,,
\end{eqnarray}
where $B(z) = z^2 + 3 \log z$, and $r_*$ is the Regge–Wheeler tortoise coordinate in $AdS_5$, one obtains a Schrödinger-like equation:
\begin{equation} \label{eqSscalar}
\partial_{r_*}^2 \psi + m_n^2 \psi = V \psi \,,
\end{equation}
where the potential $V$ is given by:
\begin{equation} \label{potential1}
V(r_*) = e^{B/2} \, \partial_{r_*}^2 e^{-B/2} \,.
\end{equation}

The above potential can be explicitly expressed as a function of the coordinate $z$ as follows:
\begin{equation}\label{potential2} V(z) = 2 + \frac{15}{4z^2} + z^2 - \pi^4 T^4 \left( \frac{3}{2} z^2 + 2 z^6 \right) + \pi^8 T^8 \left( z^{10} - \frac{9}{4} z^6 - 2 z^8 \right) ,. \end{equation} Furthermore, near the boundary and at low temperatures, the function $z(r_)$ can be approximated as
\begin{equation} z = -r_{*}\left[ 1 - \frac{r_{*}^4 \pi^4 T^4}{5} \right] \,. \end{equation} Using this approximation, the effective potential~(\ref{potential2}), expanded up to fourth order in temperature, takes the form:
\begin{equation}\label{potential3} V(r_{*}) = 2 + \frac{15}{4 r_{*}^2} + r_{*}^2 - \frac{12}{5} \pi^4 r_{*}^6 T^4\,. \end{equation}
It is observed that the thermal correction to the effective potential introduces a barrier in the infrared (IR) region   of the tortoise coordinate. Meanwhile, the oscillator-like term proportional to $r_{*}^2$, the constant term 2, and the infinite barrier near the boundary, represented by the term $15/(4r_{*}^2)$, remain unchanged from the zero-temperature case.
\begin{figure}[!htb]
	\centering
	\includegraphics[scale=0.60]{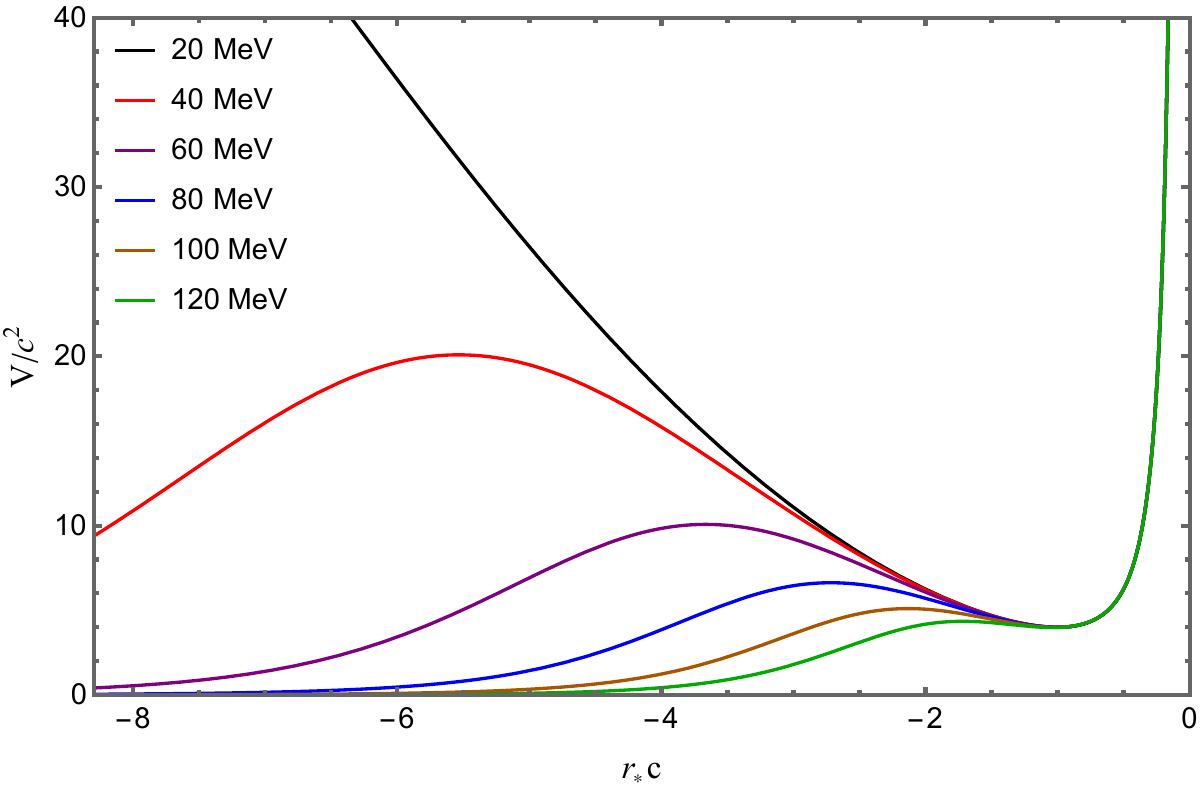}
	\caption{ The effective potential (\ref{potential1}) for the scalar glueball as a function of the tortoise coordinate $r_*$ at various temperatures.}
    \label{fig:1}
\end{figure}

In Fig.~\ref{fig:1}, we plot the effective potential as a function of the tortoise coordinate $r_*$ for various temperatures. At zero temperature, the potential features only a well. However, at nonzero temperatures, a finite barrier emerges in the IR region. In this scenario, quantum tunneling through the barrier becomes possible. If the tunneling probability is small, the system can be interpreted as a metastable state with a finite lifetime, decaying via barrier penetration. The corresponding energy becomes complex: the real part defines the energy level and governs the usual stationary behavior over time, while the imaginary part determines the decay width and induces damping. Finally, as the temperature increases, the barrier shifts toward the UV region and gradually decreases until it ultimately vanishes.

\subsection{Boundary Conditions}

Near the horizon, the asymptotic solutions of the equation consist of an incoming and an outgoing wave, given by:
\begin{eqnarray}\label{bhor}
  \psi(z) = e^{\mp i \omega r_{*}}\left(a_0+a_1\left(1-\frac{z}{z_h}\right)+a_2\left(1-\frac{z}{z_h}\right)^2+\cdots + a_n\left(1-\frac{z}{z_h}\right)^n\right)\,,
\end{eqnarray}
where the coefficients $a_0$, $a_1$, $a_2$ and $a_n$  are determined  by substituting this solution in the equation of motion and  imposing the regularity condition at horizon. On the other hand, the Schrödinger-like equation (\ref{eqSscalar})  exhibit two particulars solutions near the boundary $z=0$:
\begin{eqnarray}\label{bb}
  &&  \psi_a(z)=z^{5/2}\left(c_0+c_1 z^2+c_{2}z^4 + \cdots + c_n z^{2n}\right)  \\ \cr && \psi_b(z)=z^{-3/2}(d_0+d_{1}z^2+d_{2}z^4 + \cdots + d_n z^{2n} ) + b \, \psi_a \ln(c^2 z^2).
\end{eqnarray}
where the ellipses denote higher powers of $z$, and $d_n$, $c_n$, and $b$ are parameters that depend on $T$ and $\omega$. The wave function $\psi_a$ is normalizable due to its asymptotic behavior, $z^{5/2}$, while the wave function $\psi_b$ is non-normalizable, since it behaves asymptotically as $z^{-3/2}$.
In the case of quasinormal modes in $AdS_5$ space, two boundary conditions are imposed: an ingoing wave near the horizon and a Dirichlet condition at the boundary. Thus, the non-normalizable and outgoing wave solutions are excluded to satisfy the quasinormal mode boundary conditions.

\section{WKB approximation for quasi-stationary states}

In this section, we employ the WKB approximation to derive the generalized Bohr–Sommerfeld quantization rule for computing the quasinormal modes of a scalar field in the soft-wall model. This derivation is based on the formalism developed in \cite{Popov1968QuasiclassicalAF,Popov1991QuantizationRW,Popov1991QUANTIZATIONRF,Mur1993QuantizationRF,Mur1993TheWM}, originally introduced in the context of quantum and atomic physics.
Our analysis reveals three temperature-dependent regimes that govern the behavior of the quasinormal modes. At very low temperatures, the modes are confined near the minimum of the effective potential. As the temperature increases, the potential barrier decreases, causing the modes to shift toward the top of the potential barrier. Ultimately, the modes become localized in the above-barrier region.

\subsection{Quasinormal modes trapped at  the minimum potential}

\begin{figure}[!htb]
	\centering
	\includegraphics[scale=0.6]{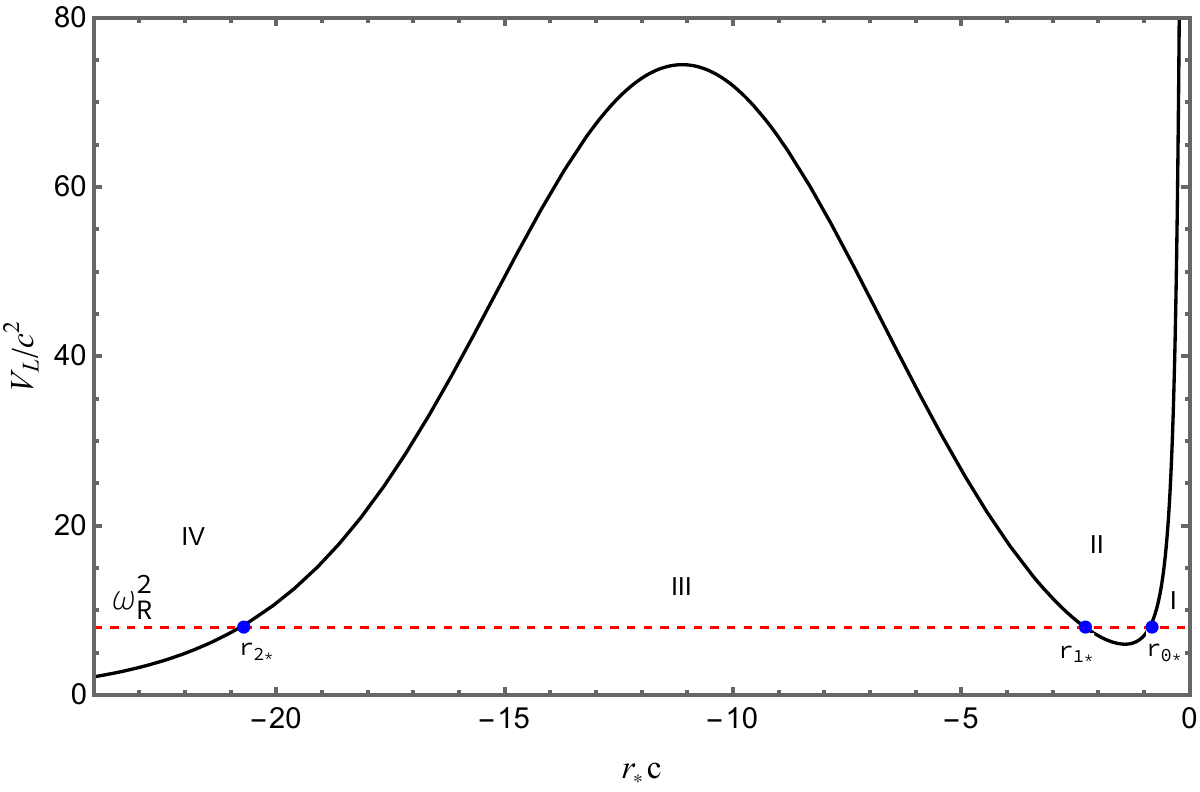}
	\caption{ The effective potential (\ref{eq:s3}) is shown as a function of the  tortoise coordinate $r_*$ at  $T=20$ $MeV$. The red dashed line denotes the real part of the frequency squared, with $\omega_R = 2.82029$ GeV.}
    \label{TurningPointsUnder}
\end{figure}

We start by analyzing the quasinormal modes with frequencies $\omega_n$ that are trapped near the minimum of the potential, corresponding to the very low-temperature case. As shown in Fig. \ref{TurningPointsUnder}, for a given real part of the frequency $\omega_R^2$, the horizontal line lies in the lower region of the potential and intersects it at the turning points $r_{0*}$, $r_{1*}$, and $r_{2*}$. There exist two classically allowed regions, II $= [r_{0*},r_{1*}]$, IV  $= ]r_{2*},r_{h*}[$ and two classically forbidden regions, I $= ]0,r_{0*}[$ and III $= [r_{1*},r_{2*}]$.

According to \cite{Karnakov_2013,scrucca2025}, the next step involves matching the WKB wave functions across the different regions. However, special care must be taken when performing this matching due to the presence of an infinite barrier near the boundary, $r_*\rightarrow 0$,  which behaves similarly to a centrifugal term in the potential as encountered in quantum mechanics. This infinite barrier introduces inaccuracies in the spectrum and affects the behavior of the wave function near the origin. Indeed, a closer examination reveals that the WKB approximation fails in the vicinity of the origin.

The first proposal to address the centrifugal problem in quantum mechanics was made by Langer in 1937 \cite{PhysRev.51.669}, during his WKB analysis of the radial wave equation. Langer observed inaccuracies in the behavior of the WKB wave function near the origin. To address this issue, a coordinate transformation was introduced:
\begin{equation}\label{coordT} r = e^x, \qquad \psi(r) = \chi(e^x) ,. \end{equation}
In this way, the origin is relocated to $-\infty$, and the matching of the WKB wave function can then be performed. As a consequence, after returning to the original coordinate, the potential is modified and becomes:
\begin{equation}\label{coordT} V_L(r) = V(r) + \frac{1}{4r^2} ,. \end{equation}
Hence, the original potential is modified by the addition of the term $1/(4r^2)$. This term leads to the correct spectrum and ensures the proper behavior of the WKB wave function at the origin. A more detailed discussion of the Langer transformation can be found in Refs.~\cite{Koike_2009,Karnakov_2013,scrucca2025}.
An alternative approach to matching the WKB solutions involves considering the exact solution in the non-quasiclassical region. In our case, for example, the solution can be obtained by analyzing the asymptotic behavior of the potential near the boundary.

Therefore, for our analysis, it is necessary to implement a Langer-type transformation due to the presence of an infinite barrier near the boundary in the holographic potential. We also require that the WKB wave function vanishes at the origin and consists solely of an ingoing wave at the horizon, ensuring the correct boundary conditions for quasinormal modes in $AdS_5$ space, as detailed in Appendix~\ref{BSHB}. The WKB wave functions are given by:
\begin{align}\label{WKBwave1}
    \left\{ \begin{array}{llll} \Psi(r_*)=\frac{\sqrt{i}C}{2\sqrt{p_L(r_*)}}\exp\left[\frac{i}{\hbar}\int_{r_{0*}}^{r_{*}}  p_L(r_*') dr_*'  \right] \,\,\,\,\,\,\, r_*>r_{0*};  \\\\  \Psi(r_*)=\frac{C}{\sqrt{p_L(r_*)}}\sin\left[\frac{1}{\hbar}\int_{r}^{r_{1*}}  p_L(r_*') dr_*' +\frac{\pi}{4} \right] \,\,\,\,\,\,\, r_{1*}<r_*<r_{0*}; \\\\ \Psi(r_*)=\frac{C_1}{\sqrt{p_L(r)}}\exp\left[-\frac{i}{\hbar}\int_{r_{2*}}^{r_*}  p_L(r_*') dr_*'  \right]+\frac{i C_1}{2\sqrt{p_L(r_*)}}\exp\left[\frac{i}{\hbar}\int_{r_{2*}}^{r_*}  p_L(r_*') dr_*'  \right]  \,\,\,\,\,\,\, r_{2*}<r_*<r_{1*};  \\\\ \Psi(r_*)=\frac{C_1}{\sqrt{p_L(r_*)}}\exp\left[\frac{i}{\hbar}\int_{r_*}^{r_{2*}}  p_L(r_*') dr_*'  \right] \,\,\,\,\,\,\, r_{h*}<r_*<r_{2*};  \end{array} \right.
\end{align}
where $p_L=\varepsilon_n-V_L(r_*)$ is the momentum of the particle and $r_*$ is the tortoise coordinate and the new potential reads
\begin{equation}\label{eq:s3}
    V_L(r_*)=V(r_*)+\frac{1}{4r_*^2}\,.
\end{equation}
with the potential $V(r_*)$  determined by  Eq. (\ref{potential1}).  Then, by matching the WKB solutions, we arrive at the following quantization condition:
\begin{equation}\label{BSqss}
    \int_{r_{1*}}^{r_{0*}}\sqrt{\varepsilon_n-V_L(r_*)}dr_{*}=\pi \left( n+\frac{1}{2}\right)-\frac{i}{4}exp\left[\frac{i}{\hbar}\int_{r_{2*}}^{r_{1*}}  p_L(r_{*}') dr_{*}'  \right]\,.
\end{equation}
Furthermore, one can expand the square root on the left-hand side of Eq.~(\ref{BSqss}) in powers of the small quantity $\varepsilon_I$, and neglect $\varepsilon_I$ on the right-hand side of the equation. This reduces the problem to solving two simpler equations:
\begin{eqnarray}
    &&\int_{r_{1*}}^{r_{0*}}\sqrt{\varepsilon_R-V_L(r_*)}dr_*=\pi \left( n+\frac{1}{2}\right) \label{eq:primeira} \\ \cr  && \varepsilon_{I}=-\frac{1}{2} \frac{e^{{2i \int_{r_{2*}}^{r_{1*}}\sqrt{\varepsilon_R-V_L(r_*)}dr_*}}}{\int_{r_{1_*}}^{r_{0_*}}\frac{1}{\sqrt{\varepsilon_{R}-V_L(r_*)}}dr_*} \label{eq:segunda}\,.
    \end{eqnarray}
The first of the above equations corresponds to the classical Bohr-Sommerfeld quantization rule for bound states, which fully determines the real part, $\varepsilon_R$. The second equation is the well-known Gamow formula, where the imaginary part, $\varepsilon_I$, is calculated directly by substituting the previously determined value of $\varepsilon_R$. 

Equations (\ref{eq:primeira}) and (\ref{eq:segunda}) provide all the information required to determine the quasinormal modes trapped in the potential minimum region. The procedure begins by substituting the potential expression from Eq.~(\ref{eq:s3}) into Eq.~(\ref{eq:primeira}) and solving it numerically to determine the real part of the frequency. Once the value of $\varepsilon_R$ is obtained, it is inserted into the Gamow formula (\ref{eq:segunda}), where the imaginary part is calculated by numerically evaluating the integral. Finally, the quasinormal frequency is determined by taking the square root: $\omega_n = \sqrt{\varepsilon_n} = \sqrt{\varepsilon_R - i\varepsilon_I}$.

\subsubsection{Results} 

For comparison, we compute the quasinormal frequencies using the shooting method \cite{Kaminski:2009ce} (see Appendix~\ref{SM} for details). The discrepancy between the two approaches is quantified by defining the deviation in the real part of the frequency as follows:
\begin{equation}\label{deviationR} \Delta_{R} = \left|\frac{\omega_{R} - \omega_{R}^{\text{sm}}}{\omega_{R}^{\text{sm}}}\right| \times 100\% \end{equation} and similarly for the imaginary part: \begin{equation}\label{deviationIm} \Delta_{I} = \left|\frac{\omega_{I} - \omega_{I}^{\text{sm}}}{\omega_{I}^{\text{sm}}}\right| \times 100\% \end{equation} where $\omega_{R/I}^{\text{sm}}$ are the quasinormal mode frequencies obtained via the shooting method, and $\omega_{R/I}$ are the frequencies computed using the WKB approximation.

Table \ref{tbl:wasss} presents the ground-state frequencies as a function of temperature, computed using Eqs.~(\ref{eq:primeira}) and (\ref{eq:segunda}), and compares them with the values obtained via the shooting method. We also present the corresponding deviation values between the two methods. At very low temperatures, the real parts of the frequencies from the two approaches show good agreement, as illustrated by the small deviations presented in the table. Nevertheless, as the temperature increases, deviations become more pronounced due to thermal effects that are not fully accounted for in the WKB approximation. Furthermore, as indicated by the deviation values, the imaginary part of the frequency calculated using the WKB approximation is systematically less accurate than its real counterpart, regardless of temperature.

\begin{table}[htbp]
  \centering
  \begin{tabular}{||c||c|c||c|c||c|c||}
    \hline
    \multicolumn{1}{||c||}{n=0} 
    & \multicolumn{2}{|c||}{Shooting Method} 
    & \multicolumn{2}{|c||}{0th Order -- BS Rule} 
    & \multicolumn{2}{|c||}{Deviation (\%)} \\
    \cline{2-7}
    \multicolumn{1}{||c||}{T (MeV)} 
    & $\omega_R$ (GeV) & $\omega_I$ (GeV) 
    & $\omega_R$ (GeV) & $\omega_I$ (GeV) 
    & $\Delta_R$ & $\Delta_I$ \\
    \hline
    $20$ & $2.82803$ & $1.63944 \times 10^{-88}$ & $2.82809$ & $1.48014 \times 10^{-88}$ & $2.1 \times 10^{-3}$ & $9.72$ \\
    $30$ & $2.82640$ & $9.51912 \times 10^{-34}$ & $2.82676$ & $8.51118 \times 10^{-34}$ & $1.28 \times 10^{-2}$ & $10.6$ \\
    $40$ & $2.82194$ & $8.29163 \times 10^{-16}$ & $2.82311$ & $7.36371 \times 10^{-16}$ & $4.14\times 10^{-2}$ & $11.2$ \\
    $50$ & $2.81203$ & $3.61824 \times 10^{-8}$  & $2.81512$ & $3.22443 \times 10^{-8}$  & $1.10 \times 10^{-1}$ & $10.9$ \\
    \hline
  \end{tabular}
  \caption{The ground state frequencies, calculated using both the shooting method and the BS rule  (\ref{eq:primeira}) and Gamow  (\ref{eq:segunda}) equations, are presented for various temperatures.}
  \label{tbl:wasss}
\end{table}

\begin{table}[htbp]
  \centering
  \begin{tabular}{||c||c|c||c|c||c|c||}
    \hline
    \multicolumn{1}{||c||}{$n=2$} 
    & \multicolumn{2}{|c||}{Shooting Method} 
    & \multicolumn{2}{|c||}{0th Order -- BS Rule} 
    & \multicolumn{2}{|c||}{Deviation (\%)} \\
    \cline{2-7}
    \multicolumn{1}{||c||}{T (MeV)} 
    & $\omega_R$ (GeV) & $\omega_I$ (GeV)
    & $\omega_R$ (GeV) & $\omega_I$  (GeV)
    & $\Delta_R$ & $\Delta_I$ \\
    \hline
    $20$ & $3.99605$ & $4.73338 \times 10^{-73}$ & $3.99616$ & $4.56426 \times 10^{-73}$ & $2.80 \times 10^{-3} $ & $3.58$ \\
    $25$ & $3.99026$ & $1.84046 \times 10^{-39}$ & $3.99055$ & $1.77003 \times 10^{-39}$ & $7.30 \times 10^{-3} $ & $3.83$ \\
    $30$ & $3.97945$ & $3.02732 \times 10^{-22}$ & $3.98008$ & $2.90906 \times 10^{-22}$ & $1.58 \times 10^{-2}$ & $3.91$ \\
    $35$ & $3.96071$ & $1.38336 \times 10^{-12}$ & $3.96201$ & $1.33389 \times 10^{-12}$ & $3.29 \times 10^{-2}$ & $3.58$ \\
    \hline \hline
    \multicolumn{1}{||c||}{$n=3$} 
    & \multicolumn{2}{|c||}{Shooting Method} 
    & \multicolumn{2}{|c||}{0th Order -- BS Rule} 
    & \multicolumn{2}{|c||}{Deviation (\%)} \\
    \cline{2-7}
    \multicolumn{1}{||c||}{T (MeV)} 
    & $\omega_R$ (GeV) & $\omega_I$  (GeV)
    & $\omega_R$ (GeV) & $\omega_I$ (GeV)
    & $\Delta_R$  & $\Delta_I$ \\
    \hline
    $20$ & $4.46453$ & $1.58433 \times 10^{-66}$ & $4.46467$ & $1.53904 \times 10^{-66}$ & $3.10 \times 10^{-3}$ & $2.86$ \\
    $25$ & $4.45328$ & $5.71864 \times 10^{-34}$ & $4.45362$ & $5.54650 \times 10^{-34}$ & $7.60 \times 10^{-3} $ & $3.01$ \\
    $30$ & $4.43184$ & $1.45494 \times 10^{-17}$ & $4.43268$ & $1.41342 \times 10^{-17}$ & $1.89 \times 10^{-2} $ & $2.85$ \\
    $35$ & $4.39306$ & $1.26223 \times 10^{-8}$ & $4.39485$ & $1.24035 \times 10^{-8}$ & $4.07 \times 10^{-2}$ & $1.73$ \\
    \hline
  \end{tabular}
  \caption{The comparison of frequencies for the  $n=2$ and  $n=3$ modes, computed using the shooting method and the BS rule  (\ref{eq:primeira}) and Gamow  (\ref{eq:segunda}) equations, across various temperatures.}
  \label{exci}
\end{table}

In Table \ref{exci}, we  present the quasinormal frequencies for the modes  $n=2$ and $n=3$, together with their respective deviations. As the excitation level increases, the results from the WKB approximation align more closely with those obtained using the shooting method, with deviations smaller than those observed for the ground state. 

\subsubsection{Higher order corrections}

Although the WKB approximation shows good agreement with the shooting method, the results can still be improved by incorporating corrections to the approximation. Let us point out that the Langer transformation introduces corrections to the wave function near the boundary, which subsequently affect the spectrum at zero temperature (see Appendix~\ref{BSzeroT}).
In fact, for the soft-wall model, the spectrum at zero temperature is exact. However, at finite temperature, although the Langer transformation still provides corrections to the wave function near the boundary, the spectrum is no longer exact. Therefore, it becomes necessary to account for additional temperature-dependent corrections, since the spectrum is already exact at zero temperature.

In this paper, we consider the first-order correction. Hence, Eq.~(\ref{eq:primeira}) is replaced by the  quantization condition introduced in Ref.~\cite{PhysRev.164.171}:
\begin{equation}\label{BScorrections} \int_{r_{1*}}^{r_{0*}} \sqrt{\varepsilon_R - V_L(r_)} , dr_ = \pi \left(n + \frac{1}{2} \right) + I_1 \end{equation}
where
\begin{eqnarray} I_1 = \frac{d^2}{d\varepsilon_R^{2}} \int_{r_{1*}}^{r_{0*}} \frac{4\left(\varepsilon_R - V_L\right)^2 + \left(2V_T - 2V_L + r_* V_T'\right)\left[12\left(\varepsilon_R - V_L\right) + \left(2V_T - 2V_L + r_{*} V_T'\right)\right]}{24 r_{*}^2 \sqrt{\varepsilon_R - V_L}}  dr_{*}. \end{eqnarray}
Here, $V_L$ is the Langer potential, which incorporates the Langer transformation, while $V_T$ denotes the potential without the $r_{*}^{-2}$ term. In other words, $V_T$ does not have an infinite barrier at $r_{*}= 0$.

For the case of the Gamow formula, equation (\ref{eq:segunda}) can be rewritten as
\begin{eqnarray}
     \varepsilon_{I}=-\frac{1}{2} \frac{e^{{2i \int_{r_{2*}}^{r_{1*}}\sqrt{\varepsilon_{R}-V(r_*)}dr_*+2\bar{I_1}}}}{\int_{r_{1*}}^{r_{0*}}\frac{1}{\sqrt{\varepsilon_{R}-V(r_*)}}dr_*} \label{eq:segunda2}\,.
\end{eqnarray}
where 
\begin{eqnarray}
\bar{I}_1=- \frac{d^2}{d\varepsilon_R^{2}}\int_{r_{1*}}^{r_{0*}} \frac{4\left(\varepsilon_R-V_{L}\right)^2 +\left(2V_T-2V_L+r_*V_T'\right) \left[12\left(\varepsilon_R-V_{L}\right)+\left(2V_T-2V_L+r_*V_T'\right) \right]}{24r^2_{*}\sqrt{V_L-\varepsilon_R}}dr_*.
\end{eqnarray}\\
The results are presented in Table \ref{tblcorrectgs}. It is evident that incorporating corrections into the BS rule  for the ground state  improves the agreement with the results obtained from the shooting method.  In fact, the deviation values decrease  when the first-order correction is taken into account. 
\begin{table}[htbp]
  \centering
  \begin{tabular}{||c||c|c||c|c||c|c||}
    \hline
    \multicolumn{1}{||c||}{$n=0$} 
    & \multicolumn{2}{|c||}{Shooting Method} 
    & \multicolumn{2}{|c||}{1st Order -- BS Rule} 
    & \multicolumn{2}{|c||}{Deviation (\%)} \\
    \cline{2-7}
    \multicolumn{1}{||c||}{T (MeV)} 
    & $\omega_R$ (GeV) & $\omega_I$ (GeV)
    & $\omega_R$ (GeV) & $\omega_I$ (GeV) 
    & $\Delta_R$ & $\Delta_I$ \\
    \hline
    $20$ & $2.82803$ & $1.63944 \times 10^{-88}$ & $2.82803$ & $1.65774 \times 10^{-88}$ & $8.29 \times 10^{-9}$ & $1.13$ \\
    $30$ & $2.82640$ & $9.51912 \times 10^{-34}$ & $2.82640$ & $9.62851 \times 10^{-34}$ & $1.77\times10^{-7}$ & $1.15$ \\
    $40$ & $2.82194$ & $8.29163 \times 10^{-16}$ & $2.82194$ & $8.39588 \times 10^{-16}$ & $1.68 \times 10^{-7}$ & $1.26$ \\
    $50$ & $2.81203$ & $3.61824 \times 10^{-8}$  & $2.81206$ & $3.67752 \times 10^{-8}$  & $1.10 \times 10^{-3}$ & $1.64$ \\
    \hline
  \end{tabular}
  \caption{Comparison of ground state frequencies obtained using the shooting method and the BS rule (\ref{BScorrections}) and Gamow formula (\ref{eq:segunda2}), which provide the real and imaginary parts of the frequency, respectively, for different temperatures.}
  \label{tblcorrectgs}
\end{table}
We also present in Table~\ref{tblcorrectgsexic} the results for the second and third excited states, corresponding to $n = 2$ and $n = 3$. As shown, the quasinormal frequencies computed via the WKB method exhibit improved accuracy, with significantly reduced deviation values relative to the zeroth-order result.

\begin{table}[htbp]
  \centering
  \begin{tabular}{||c||c|c||c|c||c|c||}
    \hline
    \multicolumn{1}{||c||}{$n=2$} 
    & \multicolumn{2}{|c||}{Shooting Method} 
    & \multicolumn{2}{|c||}{1st Order -- BS Rule} 
    & \multicolumn{2}{|c||}{Deviation (\%)} \\
    \cline{2-7}
    \multicolumn{1}{||c||}{T (MeV)} 
    & $\omega_R$  (GeV) & $\omega_I$ (GeV)
    & $\omega_R$ (GeV) & $\omega_I$  (GeV)
    & $\Delta_R$ & $\Delta_I$ \\
    \hline
    $20$ & $3.99605$ & $4.73338 \times 10^{-73}$ & $3.99605$ & $4.73411 \times 10^{-73}$ & $5.\times10^{-7}$ & $0.0168$ \\
    $25$ & $3.99026$ & $1.84046 \times 10^{-39}$ & $3.99026$ & $1.84104 \times 10^{-39}$ & $3.2 \times 10^{-6}$ & $0.0325$ \\
    $30$ & $3.97945$ & $3.02732 \times 10^{-22}$ & $3.97945$ & $3.02924 \times 10^{-22}$ & $1.9 \times 10^{-5}$ & $0.0640$ \\
    $35$ & $3.96071$ & $1.38336 \times 10^{-12}$ & $3.96072$ & $1.38546 \times 10^{-12}$ & $1.2 \times 10^{-3} $ & $0.152$ \\
    \hline \hline
    \multicolumn{1}{||c||}{$n=3$} 
    & \multicolumn{2}{|c||}{Shooting Method} 
    & \multicolumn{2}{|c||}{1st Order -- BS Rule} 
    & \multicolumn{2}{|c||}{Deviation (\%)} \\
    \cline{2-7}
    \multicolumn{1}{||c||}{T (MeV)} 
    & $\omega_R$ (GeV)  & $\omega_I$ (GeV) 
    & $\omega_R$ (GeV) & $\omega_I$ (GeV)
    & $\Delta_R$   & $\Delta_I$  \\
    \hline
    $20$ & $4.46453$ & $1.58433 \times 10^{-66}$ & $4.46453$ & $1.58444 \times 10^{-66}$ & $5 \times 10^{-7}$ & $0.0081$ \\
    $25$ & $4.45328$ & $5.71864 \times 10^{-34}$ & $4.45328$ & $5.72006 \times 10^{-34}$ & $4 \times 10^{-6}$ & $0.0254$ \\
    $30$ & $4.43184$ & $1.45494 \times 10^{-17}$ & $4.43184$ & $1.45596 \times 10^{-17}$ & $3.16 \times 10^{-5}$ & $0.0708$ \\
    $35$ & $4.39306$ & $1.26223 \times 10^{-8}$  & $4.39307$ & $1.26582 \times 10^{-8}$  & $3.7 \times 10^{-4}$ & $0.284$ \\
    \hline
  \end{tabular}
  \caption{The comparison of frequencies for the  $n=2$ and  $n=3$ modes, computed using the shooting method and the BS rule (\ref{BScorrections}) and Gamow formula (\ref{eq:segunda2}) , which provide the real and imaginary parts of the frequency, respectively, for different temperatures.}
  \label{tblcorrectgsexic}
\end{table}

At last, let us emphasize that in this regime, it is possible to obtain an analytic expression for the real part of the frequency by performing a temperature expansion. This is presented in Appendix \ref{BSzerverylowtemperature} for both the scalar and vector field cases.


\subsection{Quasinormal modes in the vicinity of the top of the potential barrier.}

\begin{figure}[!htb]
	\centering
	\includegraphics[scale=0.6]{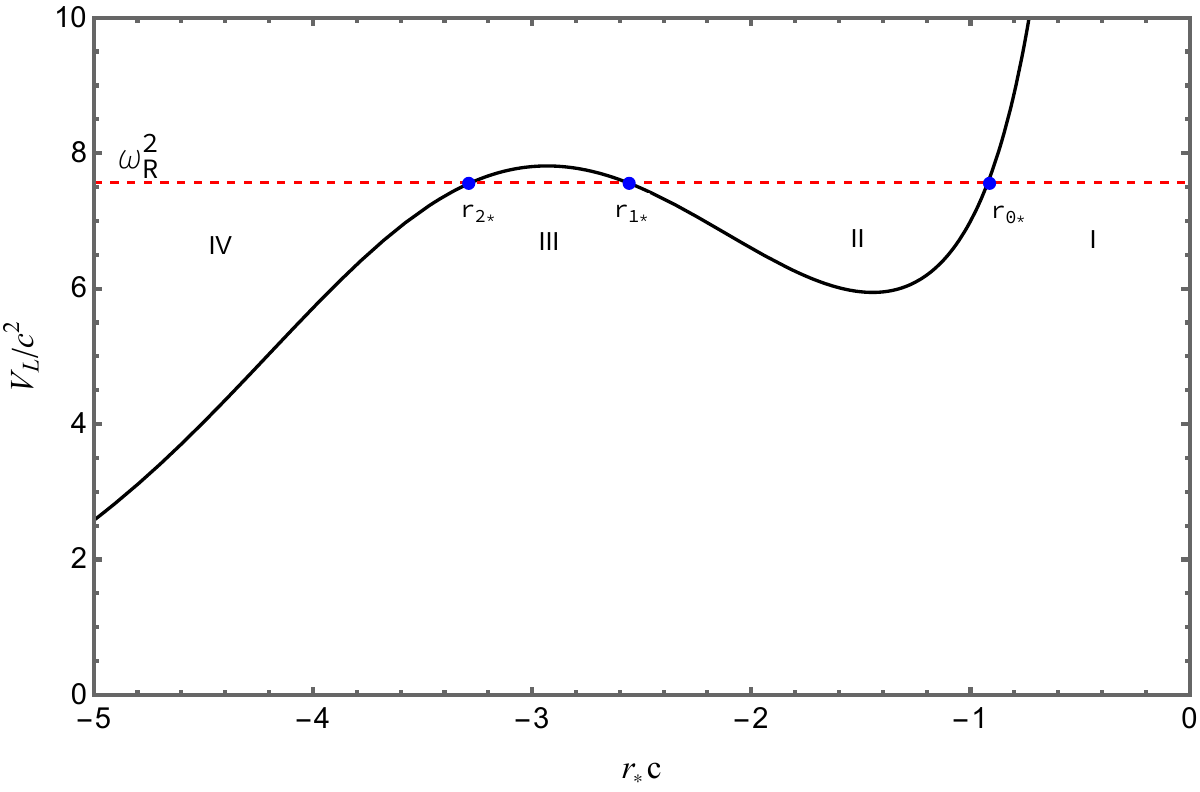}
	\caption{ The effective potential (\ref{eq:s3}) as function of  the  tortoise coordinate $r_*$ at $T=73$ $MeV$. The red dashed line indicates the value of the real part of the frequency squared, with $\omega_{R}=2.73944$ GeV. }
    \label{Near:top2}
\end{figure}

As the temperature increases, the finite barrier in the effective potential begins to decrease. Consequently, the quasinormal modes with frequency $\omega_n$, which were previously trapped at the minimum of the effective potential, gradually approach the top of the potential barrier, as illustrated in Fig.~\ref{Near:top2}.
The extension of the Bohr–Sommerfeld quantization rule to quasi-stationary states near the top of the potential barrier was developed in Refs.~\cite{Popov1968QuasiclassicalAF,Popov1991QUANTIZATIONRF,Popov1991QuantizationRW}. A more detailed discussion of this approach can be found in Ref.~\cite{Karnakov_2013}. 

The WKB wave solutions considered in this regime are given by:
\begin{align}
    \left\{ \begin{array}{llll} \Psi(r_*)=\frac{\sqrt{i}C}{2\sqrt{p_L(r_*)}}\exp\left[i\int_{r_{1*}}^{r_{0*}}  p_L(r_*') dr_*'  \right] \,\,\,\,\,\,\, r_*>r_{0*};  \\\\  \Psi(r_*)=\frac{C}{\sqrt{p_L(r_*)}}\sin\left[\int_{r_*}^{r_{1*}}  p_L(r_*') dr_*' +\frac{\pi}{4} \right] \,\,\,\,\,\,\, r_{1*}<r_*<r_{0*}; \\\\  \Psi(r_*)=\frac{C_1}{\sqrt{p_L(r_*)}}\exp\left[\frac{i}{\hbar}\int_{r_*}^{r_{2*}}  p_L(r_*') dr_*'  \right] \,\,\,\,\,\,\, r_{h*}<r_*<r_{2*};  \end{array} \right.
\end{align}
These solutions take the same form as those for quasi-stationary states in the under-barrier region. However, since the WKB approximation fails near the top of the potential barrier, the authors of Refs.~\cite{Popov1968QuasiclassicalAF,Popov1991QUANTIZATIONRF,Popov1991QuantizationRW} employed the parabolic approximation to solve the Schrödinger equation in the region $r_{2*} < r_* < r_{1*}$.
Following the same procedure, we obtain the generalized Bohr–Sommerfeld quantization rule~\cite{Popov1968QuasiclassicalAF,Popov1991QUANTIZATIONRF,Popov1991QuantizationRW}:
\begin{equation}\label{BSG2} \int_{r_{0*}}^{r_{1*}} \sqrt{\varepsilon_n - V_L(r_*)} dr_{*} = \pi \left(n + \frac{1}{2} \right) - \frac{1}{2}  \Xi(\lambda), \end{equation}
where
\begin{equation} \Xi(\lambda) = \lambda(1 - \ln \lambda) + \frac{1}{2i} \ln\left[\frac{\Gamma\left(\frac{1}{2} + i\lambda\right)}{\Gamma\left(\frac{1}{2} - i\lambda\right)\left(1 + e^{-2\pi\lambda}\right)}\right], \end{equation}
and
\begin{equation} \lambda = \frac{1}{\pi} \int_{r_{2*}}^{r_{1*}} \sqrt{V_L(r_{*}) - \varepsilon_n}  dr_{*}. \end{equation}
Here, $r_{0*}$, $r_{1*}$, and $r_{2*}$ are the turning points, and $\Gamma$ denotes the gamma function. The computation of quasinormal frequencies involves evaluating these integrals, which, in our case, must be performed numerically. Accordingly, the frequencies are obtained via numerical methods.

In Table~\ref{tablegroundstatetop1}, we present the results for the ground state computed using both the shooting method and the WKB approximation. Additionally, Table~\ref{tablegroundstatetop2} shows the corresponding results for the first three excited states, $n = 1$, $n = 2$, and $n = 3$.
As observed, the WKB approximation provides reasonable agreement with the shooting method for the real part of the frequency. However, the imaginary part of the frequency exhibits a lower level of accuracy compared to the results obtained using the shooting method.

\begin{table}[htbp]
  \centering
  \begin{tabular}{||c||c|c||c|c||c|c||}
    \hline
    \multicolumn{1}{||c||}{$n=0$} 
    & \multicolumn{2}{|c||}{Shooting Method} 
    & \multicolumn{2}{|c||}{0th Order -- BS Rule} 
    & \multicolumn{2}{|c||}{Deviation (\%)} \\
    \cline{2-7}
    \multicolumn{1}{||c||}{$T$ (MeV)} 
    & $\omega_R$ (GeV) & $\omega_I$ (GeV) 
    & $\omega_R$ (GeV) & $\omega_I$  (GeV)
    & $\Delta_R$ & $\Delta_I$ \\
    \hline
    $73$ & $2.73078$ & $1.93488 \times 10^{-2}$ & $2.73944$ & $1.71440 \times 10^{-2}$ & $3.17 \times 10^{-1}$ & $11.4$ \\
    $74$ & $2.72508$ & $2.33268 \times 10^{-2}$ & $2.73420$ & $2.07250 \times 10^{-2}$ & $3.34 \times 10^{-1}$ & $11.1$ \\
    $75$ & $2.71939$ & $2.77073 \times 10^{-2}$ & $2.72899$ & $2.46787 \times 10^{-2}$ & $3.53 \times 10^{-1} $ & $10.9$ \\
    $76$ & $2.71377$ & $3.24751 \times 10^{-2}$ & $2.72383$ & $2.89917 \times 10^{-2}$ & $3.70 \times 10^{-1}$ & $10.7$ \\
    \hline
  \end{tabular}
  \caption{The frequencies for the ground state, calculated using both the shooting method and the BS formula in the vicinity of the top of the potential barrier (\ref{BSG2}), are presented for various temperature.}
  \label{tablegroundstatetop1}
\end{table}

\begin{table}[htbp]
  \centering
  \begin{tabular}{||c||c|c||c|c||c|c||}
    \hline
    \multicolumn{1}{||c||}{$n=1$} 
    & \multicolumn{2}{|c||}{Shooting Method} 
    & \multicolumn{2}{|c||}{0th Order -- BS Rule} 
    & \multicolumn{2}{|c||}{Deviation (\%)} \\
    \cline{2-7}
    \multicolumn{1}{||c||}{$T$ (MeV)} 
    & $\omega_R$ (GeV) & $\omega_I$ (GeV)
    & $\omega_R$  (GeV) & $\omega_I$ (GeV)
    & $\Delta_R$ & $\Delta_I$ \\
    \hline
    $54$ & $3.34987$ & $5.03445 \times 10^{-3}$ & $3.35340$ & $4.71279 \times 10^{-3}$ & $1.05 \times 10^{-1} $ & $6.39$ \\
    $55$ & $3.33770$ & $8.47488 \times 10^{-4}$ & $3.34161$ & $7.96662 \times 10^{-3}$ & $1.17 \times 10^{-1} $ & $6.00$ \\
    $56$ & $3.32484$ & $1.32819 \times 10^{-3}$ & $3.32913$ & $1.25368 \times 10^{-2}$ & $1.29 \times 10^{-1} $ & $5.61$ \\
    $57$ & $3.31155$ & $1.95701 \times 10^{-3}$ & $3.31621$ & $1.85427 \times 10^{-2}$ & $1.40 \times 10^{-1} $ & $5.25$ \\
    \hline \hline
    \multicolumn{1}{||c||}{$n=2$} 
    & \multicolumn{2}{|c||}{Shooting Method} 
    & \multicolumn{2}{|c||}{0th Order -- BS Rule} 
    & \multicolumn{2}{|c||}{Deviation (\%)} \\
    \cline{2-7}
    \multicolumn{1}{||c||}{$T$ (MeV)} 
    & $\omega_R$ (GeV) & $\omega_I$ (GeV)
    & $\omega_R$ (GeV) & $\omega_I$ (GeV)
    & $\Delta_R$ & $\Delta_I$ \\
    \hline
    $45$ & $3.87173$ & $1.37768 \times 10^{-3}$ & $3.87373$ & $1.31145 \times 10^{-3}$ & $5.17 \times 10^{-2} $ & $4.80$ \\
    $46$ & $3.85442$ & $3.65457 \times 10^{-3}$ & $3.85673$ & $3.49329 \times 10^{-3}$ & $5.99  \times 10^{-2} $ & $4.41$ \\
    $47$ & $3.83488$ & $8.15134 \times 10^{-3}$ & $3.83753$ & $7.82874 \times 10^{-3}$ & $6.91  \times 10^{-2}$ & $3.96$ \\
    $48$ & $3.81373$ & $1.56110 \times 10^{-2}$ & $3.81670$ & $1.50626 \times 10^{-2}$ & $7.79  \times 10^{-2} $ & $3.51$ \\
    \hline \hline
    \multicolumn{1}{||c||}{$n=3$} 
    & \multicolumn{2}{|c||}{Shooting Method} 
    & \multicolumn{2}{|c||}{0th Order -- BS Rule} 
    & \multicolumn{2}{|c||}{Deviation (\%)} \\
    \cline{2-7}
    \multicolumn{1}{||c||}{$T$ (MeV)} 
    & $\omega_R$ (GeV) & $\omega_I$ (GeV)
    & $\omega_R$  (GeV)& $\omega_I$  (GeV)
    & $\Delta_R$   & $\Delta_I$ \\
    \hline
    $38$ & $4.35524$ & $2.57519 \times 10^{-5}$ & $4.35629$ & $2.46881 \times 10^{-5}$ & $2.42  \times 10^{-2} $ & $4.13$ \\
    $39$ & $4.33843$ & $1.80070 \times 10^{-4}$ & $4.33967$ & $1.72889 \times 10^{-4}$ & $2.86  \times 10^{-2} $ & $3.99$ \\
    $40$ & $4.31833$ & $9.40440 \times 10^{-4}$ & $4.31979$ & $9.05503 \times 10^{-4}$ & $3.38  \times 10^{-2}$ & $3.71$ \\
    $41$ & $4.29411$ & $3.61913 \times 10^{-3}$ & $4.29586$ & $3.49147 \times 10^{-3}$ & $4.07  \times 10^{-2} $ & $3.52$ \\
    \hline
  \end{tabular}
  \caption{The comparison of frequencies for the  $n=1$, $n=2$ and  $n=3$ modes, computed using the shooting method and the BS formula in the vicinity of the top of the potential barrier (\ref{BSG2}), across various temperatures.}
  \label{tablegroundstatetop2}
\end{table}

Let us point out that the generalized Bohr–Sommerfeld rule~\eqref{BSG2} reduces to the quantization condition for modes trapped near the potential minimum~\eqref{BSqss} in the limit $\lambda \rightarrow \infty$. The parameter $\lambda$ serves as an indicator of the quasinormal modes location: large values of $\lambda$ correspond to modes near the potential minimum, while small values indicate proximity to the top of the potential barrier~\cite{Karnakov_2013}.

\subsection{Quasinormal modes in the above-barrier region}

\begin{figure}[!htb]
	\centering
	\includegraphics[scale=0.6]{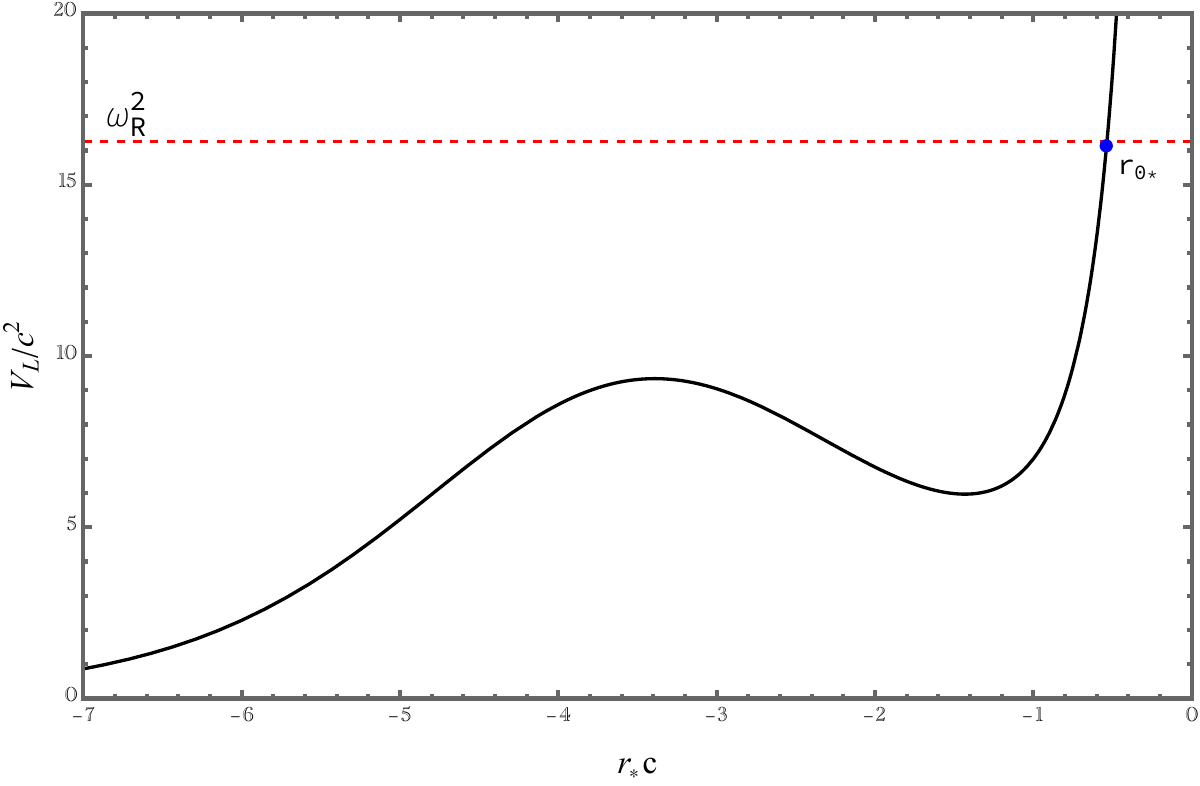}
	\caption{ The effective potential (\ref{eq:s3}) as a function of the tortoise coordinate $r_*$ at $T=64$ $MeV$. The red dashed line represents the  real part of the frequency squared, with  $\omega_R=4.0329$ GeV.}
    \label{Abovetop}
\end{figure}

Now, let us examine the case in which the modes are localized above the potential barrier, see Fig.~\ref{Abovetop}. There is a connection between the transition of the modes to the above-barrier region and the behavior of the turning points. At very low temperatures, the turning points $r_{0*}$, $r_{1*}$, and $r_{2*}$ are real. As the temperature increases, the turning point $r_{2*}$ gradually approaches $r_{1*}$ until they collide.
The temperature at which this collision occurs corresponds to the point where the quasinormal modes reach the top of the potential barrier. Above this temperature, the modes lie entirely in the above-barrier region. Furthermore, the turning points $r_{1*}$ and $r_{2*}$, which were previously real, become complex.

The localization of the modes above the barrier can be interpreted as a signal of the dissociation of the glueball particle on the gauge theory side. Within the framework of quantum mechanics and atomic physics~\cite{Karnakov_2013}, it has been emphasized that above-barrier resonance states exhibit large widths, often exceeding the spacing between neighboring states. Consequently, these states lack distinct individual properties. Hence, it is reasonable to interpret this behavior as an indication of particle dissociation in the dual field theory.
\begin{figure}[!htb]
	\centering
	\includegraphics[scale=0.6]{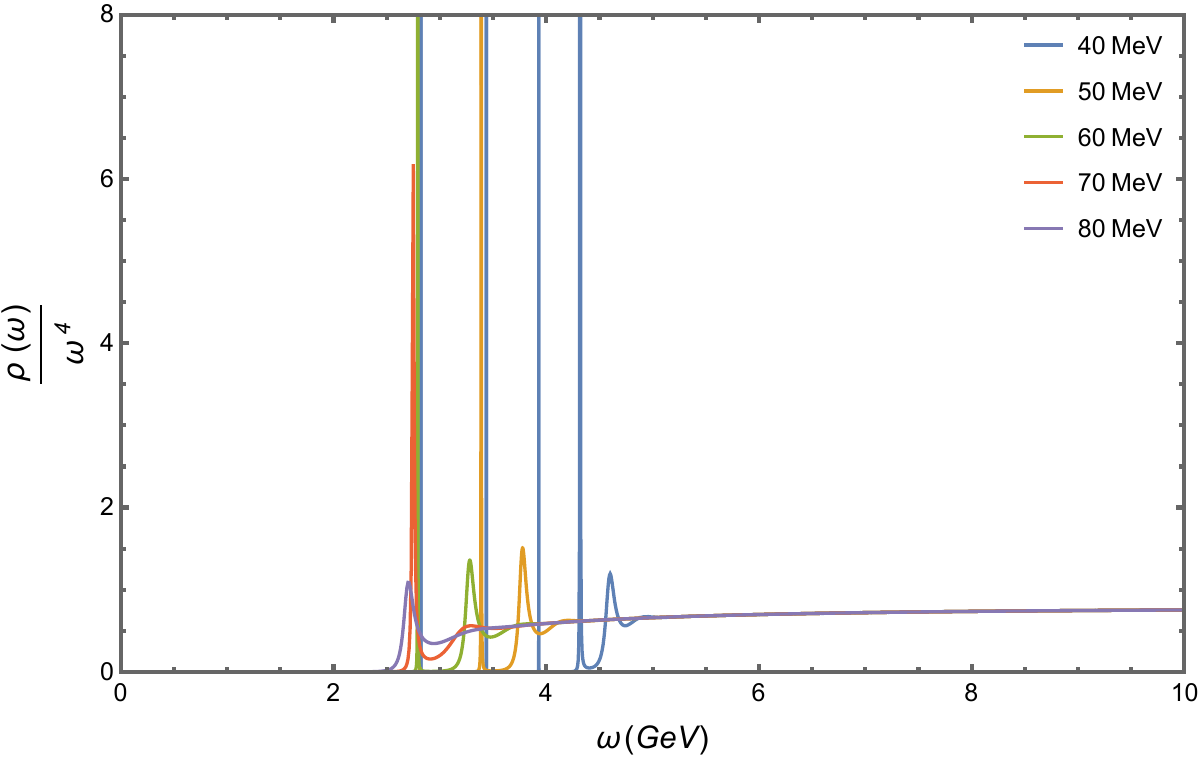}
	\caption{ The spectral function of the scalar glueballs at various temperatures.}
    \label{Spectral Function}
\end{figure}

To illustrate this behavior, we plot the spectral function at different temperatures in Fig.~\ref{Spectral Function}. The figure shows that the peak of the spectral function\footnote{The spectral function of the scalar field is derived in Appendix~\ref{SpectralFunction}.} decreases significantly as the modes move into the above-barrier region. Specifically, the temperatures at which the modes reach the top of the potential barrier are approximately 76.9 MeV for the ground state, 57.7 MeV for the first excitation, 48.5 MeV for the second excitation, and 42.7 MeV for the third excitation.

Although the turning points $r_{1*}$ and $r_{2*}$ shift into the complex plane, the WKB approximation remains applicable as $\lambda$ increases, indicating that the quasi-stationary states move away from the vicinity of the top of the barrier. This extension has been thoroughly studied in Refs.~\cite{Mur1993QuantizationRF, Mur1993TheWM}.
The generalized Bohr–Sommerfeld quantization rule~(\ref{BSG2}) for quasinormal modes above the potential barrier can be obtained through analytical continuation, implemented via the formal substitution~\cite{Mur1993TheWM, Karnakov_2013}: \begin{equation}\label{analyticcontinuation} \lambda \rightarrow \lambda e^{-2\pi i}, \end{equation} applied to the quantization rule~(\ref{BSG2}). The limit $\left| \lambda \right| \gg 1$ is considered to ensure the validity of the WKB approximation. Consequently, we do not consider solutions located very close to the top of the barrier.
Under this assumption, Eq.~(\ref{BSG2}) takes the following form~\cite{Mur1993QuantizationRF, Karnakov_2013}:
\begin{equation}\label{BSGAC}
    \oint_{C´}\sqrt{\varepsilon_n-V_L(r_*)}dr_*=\int_{r_{2*}}^{r_{0*}}\sqrt{\varepsilon_{n}-V_L(r_*)}dr_{*}=\pi\left(n+\frac{1}{2}\right)\,.
\end{equation}
where the integration  contour  $C´$ encloses the turning points $r_{0*}$ and $r_{2*}$. In the context of quantum mechanics, this formula was applied in Ref.~\cite{PhysRevA.37.4079} to compute the resonances of the anharmonic oscillator.

It can be shown that WKB corrections can still be incorporated to achieve more accurate frequency calculations~\cite{Mur1993QuantizationRF,Mur1993TheWM}. Here, we take into account the first-order correction. Thus, Eq.~(\ref{BSGAC}) can be rewritten as:
\begin{equation}\label{BSGACCorrections}
     \int_{r_{2*}}^{r_{0*}}\sqrt{\varepsilon_n-V_L(r_*)}dr_*=\pi \left( n+\frac{1}{2}\right)+I_1\,,
\end{equation}
where
\begin{eqnarray}
I_1= \frac{d^2}{d\varepsilon_n^{2}}\int_{r_{2*}}^{r_{0*}} \frac{4\left(\varepsilon_n-V_{L}\right)^2 +\left(2V_T-2V_L+r_*V_T'\right) \left[12\left(\varepsilon_n-V_{L}\right)+\left(2V_T-2V_L+r_*V_T'\right) \right]}{24r^2_{*}\sqrt{\varepsilon_n-V_L}}dr_*\,.
\end{eqnarray}

\begin{table}[htbp]
  \centering
  \begin{tabular}{||c||c|c||c|c||c|c||}
    \hline
    \multicolumn{1}{||c||}{$n=1$} 
    & \multicolumn{2}{|c||}{Shooting Method} 
    & \multicolumn{2}{|c||}{1st Order -- BS Rule} 
    & \multicolumn{2}{|c||}{Deviation (\%)} \\
    \cline{2-7}
    $T$ (MeV) & $\omega_R$  (GeV) & $\omega_I$  (GeV) & $\omega_R$  (GeV) & $\omega_I$  (GeV) & $\Delta_R$ & $\Delta_I$ \\
    \hline
    $70$  & $3.17672$ & $0.20736$ & $3.17544$ & $0.20824$ & $4.03 \times 10^{-2}$ & $4.26 \times 10^{-1}$ \\
    $80$  & $3.15010$ & $0.40032$ & $3.15133$ & $0.40053$ & $3.90 \times 10^{-2}$ & $5.24 \times 10^{-2}$ \\
    $90$  & $3.16637$ & $0.59476$ & $3.16661$ & $0.59453$ & $7.60 \times 10^{-3}$ & $3.87 \times 10^{-2}$ \\
    $100$ & $3.20797$ & $0.78713$ & $3.20809$ & $0.78675$ & $3.70 \times 10^{-3}$ & $4.83 \times 10^{-2}$ \\
    \hline\hline
    \multicolumn{1}{||c||}{$n=2$} 
    & \multicolumn{2}{|c||}{Shooting Method} 
    & \multicolumn{2}{|c||}{1st Order -- BS Rule} 
    & \multicolumn{2}{|c||}{Deviation (\%)} \\
    \cline{2-7}
    $T$ (MeV) & $\omega_R$  (GeV) & $\omega_I$  (GeV) & $\omega_R$  (GeV)& $\omega_I$  (GeV) & $\Delta_R$ & $\Delta_I$ \\
    \hline
    $65$  & $3.61282$ & $0.43055$ & $3.61289$ & $0.43057$ & $1.94 \times 10^{-3}$ & $4.65 \times 10^{-3}$ \\
    $70$  & $3.61388$ & $0.57422$ & $3.61393$ & $0.57420$ & $1.38 \times 10^{-3}$ & $3.48 \times 10^{-3}$ \\
    $75$  & $3.63005$ & $0.71732$ & $3.63010$ & $0.71728$ & $1.38 \times 10^{-3}$ & $5.58 \times 10^{-3}$ \\
    $80$  & $3.65820$ & $0.85908$ & $3.65825$ & $0.85901$ & $1.37 \times 10^{-3}$ & $8.15 \times 10^{-3}$ \\
    \hline\hline
    \multicolumn{1}{||c||}{$n=3$} 
    & \multicolumn{2}{|c||}{Shooting Method} 
    & \multicolumn{2}{|c||}{1st Order -- BS Rule} 
    & \multicolumn{2}{|c||}{Deviation (\%)} \\
    \cline{2-7}
    $T$ (MeV) & $\omega_R$  (GeV)& $\omega_I$ (GeV) & $\omega_R$ (GeV) & $\omega_I$  (GeV) & $\Delta_R$ & $\Delta_I$ \\
    \hline
    $52$ & $4.05026$ & $0.30317$ & $4.05030$ & $0.30234$ & $9.88 \times 10^{-4}$ & $2.73 \times 10^{-1}$ \\
    $56$ & $4.02498$ & $0.45337$ & $4.02500$ & $0.45338$ & $4.97 \times 10^{-4}$
 & $2.21 \times 10^{-3}$ \\
    $60$ & $4.02058$ & $0.60528$ & $4.02061$ & $0.60528$ & $7.44 \times 10^{-4}$ & $3.59 \times 10^{-4}$ \\
    $64$ & $4.03238$ & $0.75692$ & $4.03239$ & $0.75692$ & $5.90 \times 10^{-5}$ & $1.04 \times 10^{-4}$ \\
    \hline
  \end{tabular}
  \caption{The frequencies for various states, calculated using both the shooting method and the BS formula (\ref{BSGACCorrections})   are presented across a range of temperatures.}
  \label{tbAbove}
\end{table}

In Table~\ref{tbAbove}, we present the results obtained for the modes $n = 1$, $2$, and $3$ using Eq.~(\ref{BSGACCorrections}) and the shooting method for different temperature values. The frequencies calculated using the quantization rule~(\ref{BSGACCorrections}) show good agreement with those obtained from the shooting method, with deviations remaining below $1\%$.
Moreover, the width, which was small in other regimes, increases considerably in this case, further supporting the interpretation of dissociation in this regime.

Although the results for excited quasinormal modes show strong agreement with the shooting method, additional considerations are necessary at higher temperatures and for the ground state. At certain temperatures, the maximum of the barrier and the minimum of the well in the holographic potential converge to the same value.

The criterion for determining extrema in a potential function allows us to estimate the temperature at which this transition occurs, which, in our case, is approximately $T_v \approx 101$ MeV. Beyond this temperature, we observe that the first derivative of the potential vanishes only for complex values of the tortoise coordinate, while the turning point persists and does not disappear, even after the barrier is no longer present.

This behavior suggests that the WKB approximation used here does not fully capture the transition, necessitating the use of alternative methods. Additionally, we do not observe any significant change in the behavior of quasinormal modes, in contrast to the expected shift to the above-barrier region.  This indicates that it is possible to apply the BS formula in the above-barrier region. To demonstrate this, we calculate the quasinormal frequencies for temperatures $T > 101\,\text{MeV}$ and show that the results remain consistent with those obtained using the shooting method. The results for the ground state are presented in Table~\ref{tbAbove2}. The deviation values also remain below  $1\%$.

\begin{table}[htbp]
  \centering
  \begin{tabular}{|c|c|c||c|c||c|c|}
    \hline
    \multicolumn{1}{|c|}{$n=0$} 
    & \multicolumn{2}{|c||}{Shooting Method} 
    & \multicolumn{2}{|c||}{1st Order -- BS Rule} 
    & \multicolumn{2}{|c||}{Deviation (\%)} \\
    \cline{2-7}
    $T$ (MeV) & $\omega_R$ (GeV) & $\omega_I$ (GeV) & $\omega_R$ (GeV) & $\omega_I$  (GeV) &  $\Delta_R$  & $\Delta_I$  \\
    \hline
    $140$ & $2.65980$ & $0.62334$ & $2.66092$ & $0.62140$ & $4.21 \times 10^{-2}$ & $3.11 \times 10^{-1}$ \\
    $180$ & $2.80639$ & $1.03137$ & $2.81202$ & $1.02607$ & $2.00 \times 10^{-1}$ & $5.13 \times 10^{-1}$  \\
    $220$ & $3.01322$ & $1.42228$ & $3.02195$ & $1.41580$ & $2.90 \times 10^{-1}$ & $4.55 \times 10^{-1}$ \\
    $260$ & $3.26463$ & $1.79835$ & $3.27017$ & $1.78827$ & $1.70 \times 10^{-1}$ & $5.61 \times 10^{-1}$ \\
    \hline
  \end{tabular}
  \caption{The frequencies for for the ground state, calculated using both the shooting method and the BS formula  (\ref{BSGACCorrections}), are presented across a range of temperatures.}
  \label{tbAbove2}
\end{table}

\subsection{Results at finite momentum}

Here, we present the results obtained by taking into account the momentum of the particle. In this case, the holographic potential is expressed as:
\begin{equation}\label{momentumq2}
    V(r_*)=e^{B/2}\partial_{r_{*}}^2e^{-B/2}+fq^2
\end{equation}
where $B(z)=z^2+3\log z$, $r_*$ is the tortoise coordinate and $q$ is momentum of wave plane. The holographic potential can again be written in $z$ coordinate:
\begin{equation}\label{potentialq}
    V(z)=2+q^2+\frac{15}{4z^2}+z^2-\pi^4T^4\left(\frac{3}{2}z^2+2z^6+q^4z^4\right)+\pi^8T^8\left(z^{10}-\frac{9}{4}z^6-2z^8\right)\,.
\end{equation} 
In this case, the quasinormal mode frequencies are calculated in the same way as in the zero-momentum case. The potential~(\ref{momentumq2}) is then inserted into the appropriate BS quantization rule.
\begin{figure}[!htb]
	\centering
	\includegraphics[scale=0.45]{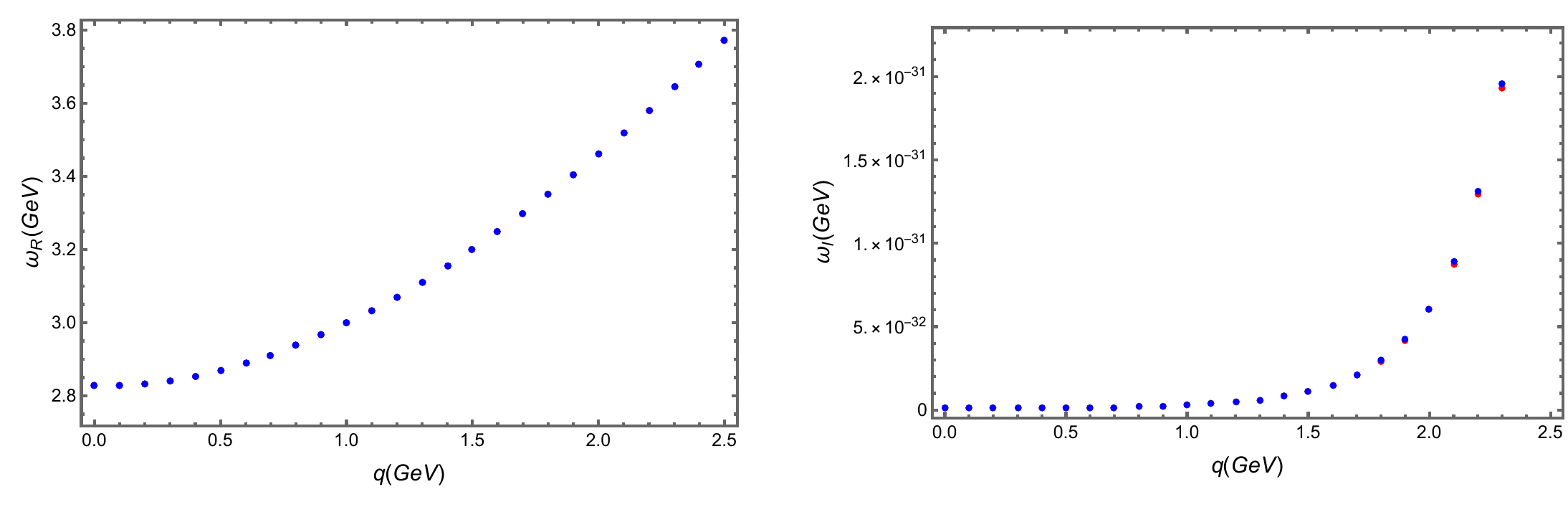}
	\caption{ The quasinormal frequencies for the ground state at $T=30$ MeV are presented as a function of the momentum $q$. The blue dots represent the values obtained using the BS quantization formula (\ref{BScorrections}) and Gamow formula (\ref{eq:segunda2}), while the red dots correspond to the frequencies computed via the shooting method.}
    \label{momentum1}
\end{figure}

\begin{figure}[!htb]
	\centering
	\includegraphics[scale=0.45]{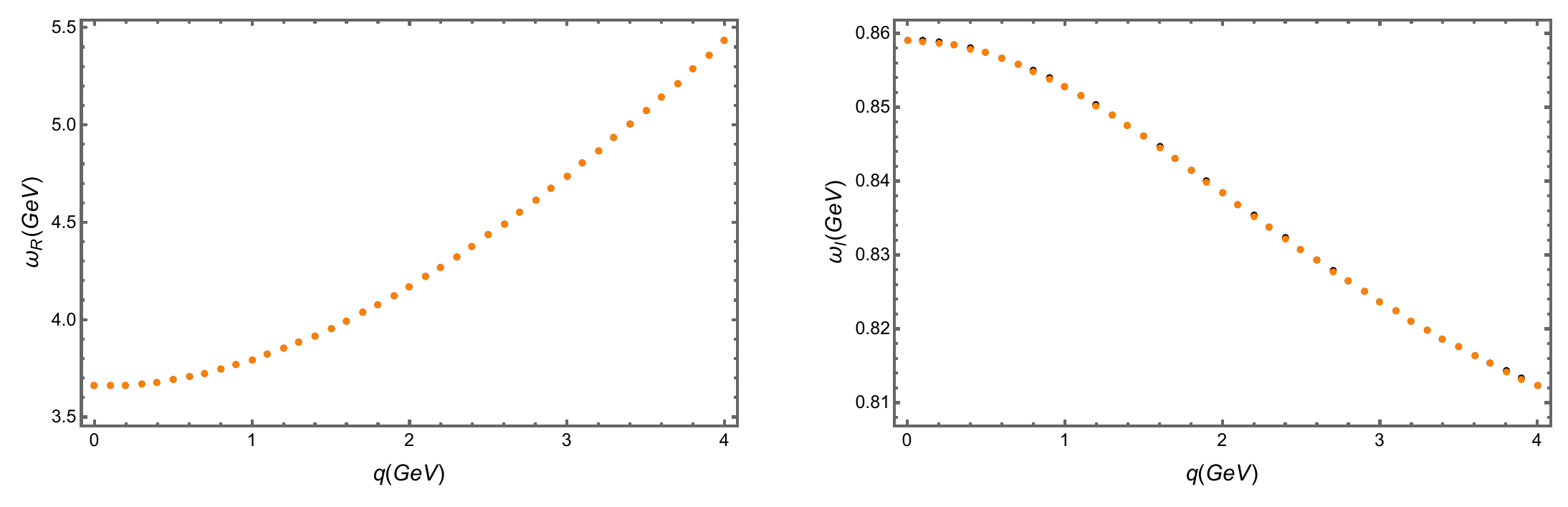 }
	\caption{ The quasinormal  frequencies for the second mode at $T=80$ MeV are presented as a function of the momentum $q$. The orange dots represent the values obtained using the BS formula   (\ref{BSGACCorrections}), while the black dots correspond to the frequencies computed via the shooting method.}
    \label{momentum2}
\end{figure}

In Fig.~\ref{momentum1}, we show the ground-state frequencies calculated using the Bohr–Sommerfeld quantization rule~(\ref{BScorrections}) and the Gamow formula~(\ref{eq:segunda2}), compared to those obtained using the shooting method for quasinormal modes in the under-barrier region. The blue dots represent the frequencies derived from the WKB approximation, while the red dots correspond to those obtained via the shooting method.
Furthermore, in Fig.~\ref{momentum2}, we present the quasinormal mode results for the $n = 2$ excitation at $T = 80$ MeV, which corresponds to a mode lying in the above-barrier region. In fact, the quasinormal modes are accurately described by the WKB approximation.

\subsection{Results higher excitations}

The WKB approximation in quantum mechanics is well known for providing increasingly accurate results as $n\rightarrow \infty $. This behavior was also observed in our analysis of the first four modes. Moreover, the WKB approximation offers an alternative method for analyzing quasinormal modes at very low temperatures with high precision, a task that is difficult to accomplish using other approaches.
\begin{figure}[!htb]
	\centering
	\includegraphics[scale=0.4]{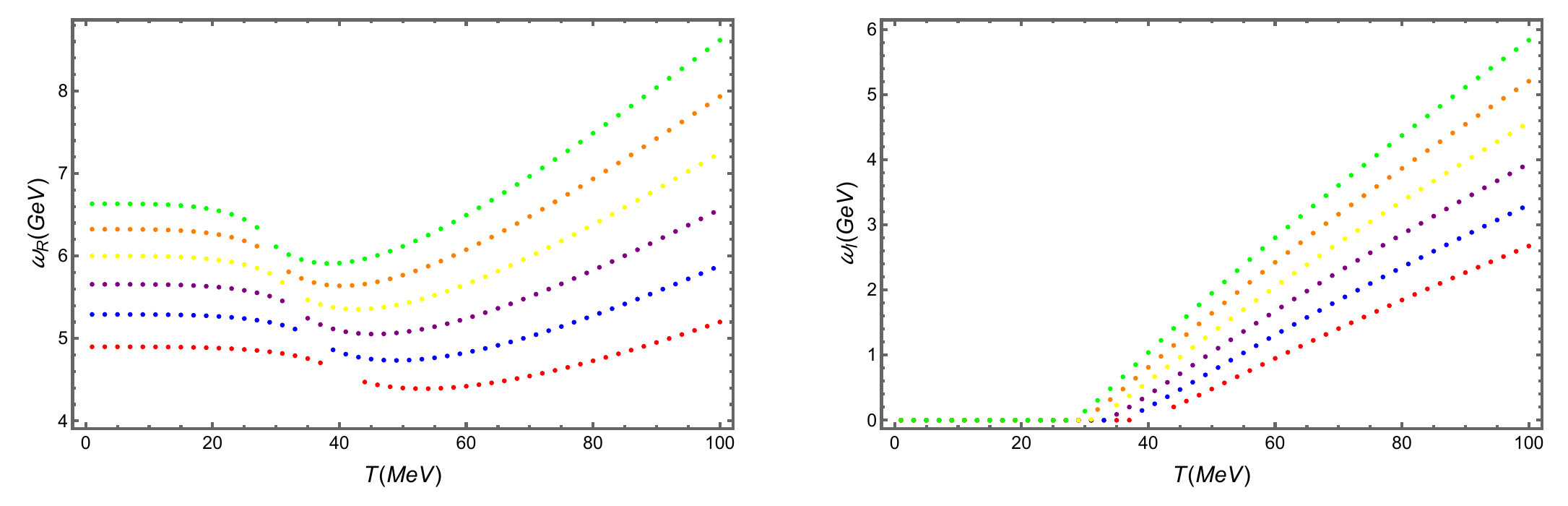}
	\caption{The quasinormal  frequencies for modes ranging from $n=4$ to $n=9$ are calculated using the WKB approximation. The $n=4$ mode is represented by red, the $n=5$ mode by blue, the $n=6$ mode by purple, and the $n=7$ mode by purple, while the $ n=8$ mode is shown in orange and the $n=9$ mode in green.  }
    \label{Higherextications}
\end{figure}

Using the shooting method, we obtained results only for temperatures up to 20 MeV. Achieving lower temperatures requires higher numerical precision. Additionally, at very low temperatures, the proximity of states can lead to mixing effects, further complicating the analysis.
Another widely used approach for studying quasinormal modes in holographic QCD models is the power series expansion method~\cite{Horowitz:1999jd,Miranda:2009uw,Mamani:2013ssa}. However, this technique is more suitable for high-temperature regimes and provides reliable results only for modes with large widths.
In contrast, the Breit–Wigner method~\cite{Berti:2009wx} is restricted to modes with small widths and is therefore applicable only in low-temperature regimes. However, it also requires greater numerical precision.

To demonstrate that the generalized Bohr–Sommerfeld quantization rule complements the study of quasinormal modes in holographic QCD models, we calculate the frequencies for excitation levels ranging from $n = 4$ to $n = 9$. The corresponding results are shown in Fig.~\ref{Higherextications}. These quasinormal modes exhibit behavior consistent with the modes discussed earlier, reinforcing the robustness of our analysis.

As expected, the dissociation process occurs rapidly and becomes more pronounced at higher excitation levels. The critical temperatures at which these modes transition beyond the barrier are approximately $T_{n=4} \approx 38$ MeV, $T_{n=5} \approx 36$ MeV, $T_{n=6} \approx 34$ MeV, $T_{n=7} \approx 32$ MeV, $T_{n=8} \approx 30$ MeV, and $T_{n=9} \approx 28$ MeV.

\section{Vector Field}

The application of the WKB approximation is not restricted to the scalar-field case; in the soft-wall model, it can also be extended to other fields in AdS space. To illustrate this, we compute the quasinormal modes of a vector field using the WKB approach.

By following the same procedure outlined for the scalar field, one can derive a Schrödinger-type equation for the vector field, with the corresponding holographic potential given in:\cite{Mamani:2013ssa,Braga:2016wkm,MartinContreras:2021bis}:
\begin{equation}\label{HolographicpV}
    V(r_*)=e^{B_V/2}\partial_{r_*}^2e^{-B_V/2}
\end{equation}
where $B_V = -\log z + \Phi$, and the tortoise coordinate $r_*$ retains the same form used previously. In the case of the soft-wall model, where $\Phi = c^2 z^2$, the effective potential is given by
\begin{equation}\label{potentialVector} V(z) = \frac{3}{4z^2} + z^2 + \pi^4 T^4 \left( \frac{z^2}{2} + 4z^4 - 2z^6 \right) + \pi^8 T^8 \left( z^{10} - \frac{5}{4} z^6 - 4z^8 \right)\,. \end{equation}
Here, we choose $c = 1$ for convenience.
\begin{figure}[!htb]
	\centering
	\includegraphics[scale=0.45]{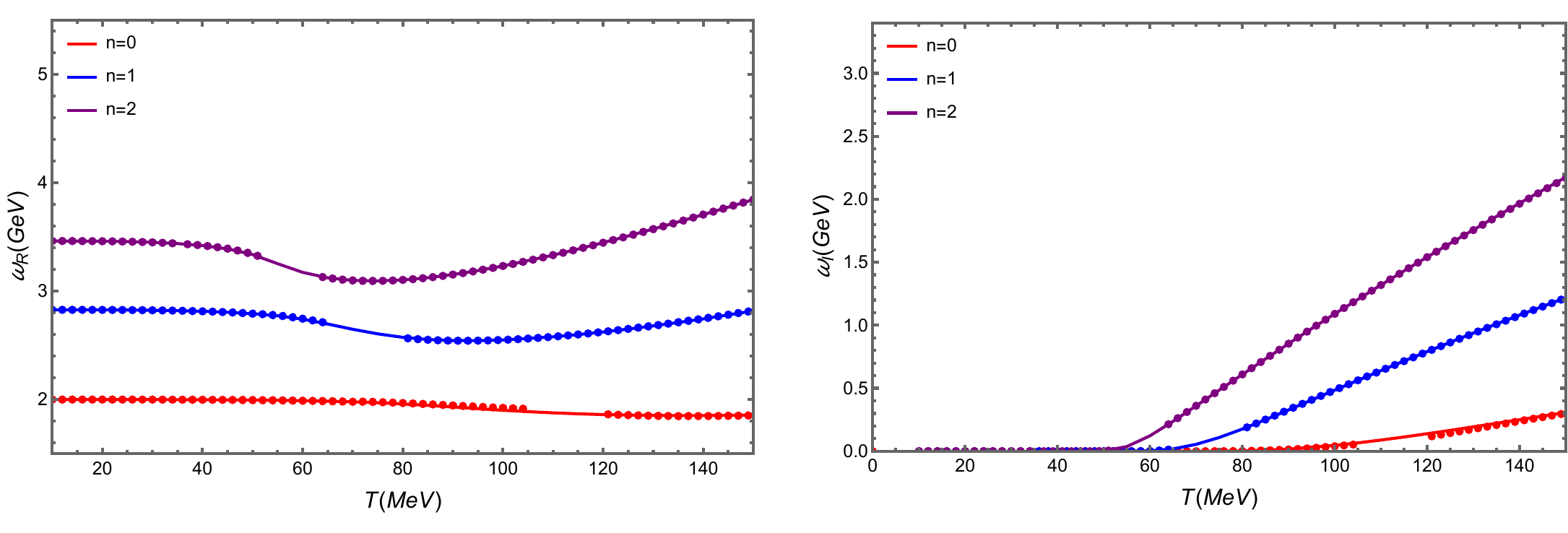}
	\caption{The frequencies for the first three modes of the vector field in the soft wall model, computed via the BS formula (\ref{BSGACCorrections}), are compared with the results obtained using the shooting method. The dot markers represent the values calculated using the WKB approximation, while the solid lines correspond to the results from the shooting method. The red color represents the $n=0$ modes, the blue color denotes $n=1$, and the purple color corresponds to the $n=2$ mode.}
    \label{VectorL1}
\end{figure}

The results for the first three modes of the original soft-wall model are displayed in Fig.~\ref{VectorL1}. The dotted points represent the frequencies calculated using the WKB approximation, while the solid lines correspond to the results obtained via the shooting method.
Notice that the WKB approximation shows good agreement with the shooting method. Furthermore, it is important to highlight that the region without dotted points arises because the above-barrier Bohr–Sommerfeld formula~(\ref{Abovetop}) cannot be applied near the top of the potential barrier due to the condition $\left| \lambda \right| \gg 1$ 

Finally, we emphasize the temperatures at which the quasinormal modes are localized near the top of the potential barrier, as shown in Table~\ref{tbAboveVnonandLt}. As in previous case, this behavior can be interpreted as an indication of particle dissociation on the gauge theory side. Moreover, the complete disappearance of the barrier occurs only at $T_v \approx 200$ MeV.

\begin{table}[htbp]
  \centering
  \begin{tabular}{||c||c||}
    \hline \hline
    $n$ & $T_{\Phi}$(MeV) \\  
    \hline
    0 & 104.7  \\  
    \hline
    1 & 67.5  \\  
    \hline
    2 & 53.9  \\  
    \hline \hline
  \end{tabular}
  \caption{Temperatures at which quasinormal modes reach the top of the potential barrier  for different modes. The values correspond to $\Phi=z^2$.}
  \label{tbAboveVnonandLt}
\end{table}

\subsection{Tangent Model}

Unlike the previously discussed holographic models, the effective potential of the heavy vector meson model proposed in Refs. \cite{Braga:2017bml,Braga:2018zlu,Braga:2018hjt} exhibits a barrier that persists at higher temperatures.  Therefore, the quasinormal modes  can be computed using the formula (\ref{BSGACCorrections}) over a wide range temperature without concerns about the collision between the maximum and minimum  local of the holographic potential, which signals the disappearance of the barrier.
\begin{figure}[!htb]
	\centering
	\includegraphics[scale=0.6]{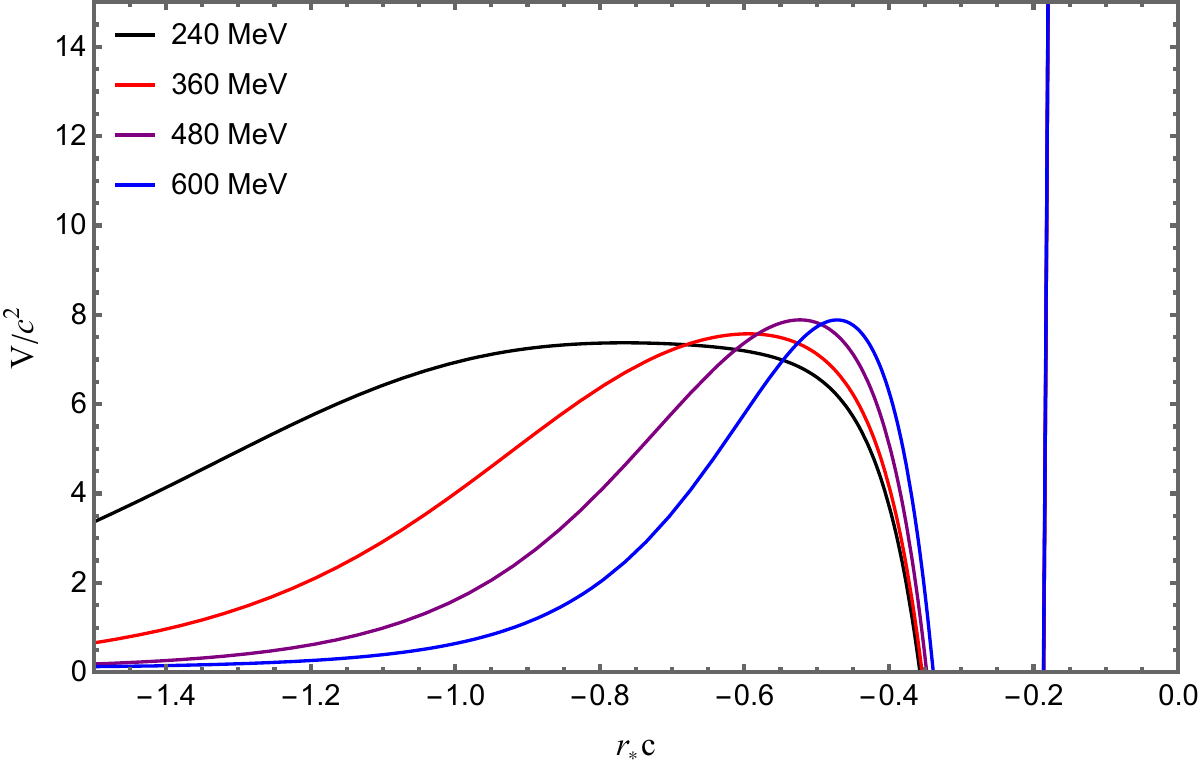}
	\caption{ The holographic potential (\ref{potential1}) as function of  the  tortoise coordinate $r_*$ at temperature $\bar{T}=0.02$ for the scalar glueball.}
    \label{fig:1Tan}
\end{figure}

The holographic potential is calculated using the same procedure described for the vector case above, following the formula (\ref{HolographicpV}). However, the dilaton is given by:
\begin{equation}
\phi(z)=k^2z^2+Mz+\tanh\left(\frac{1}{Mz}-\frac{k}{ \sqrt{\Gamma}}\right)\,.
\label{dilatonModi}
\end{equation}
where the parameter $k$ represents the quark mass, $\Gamma$ the string tension of the strong quark anti-quark interaction and $M$ is a mass scale associated with non hadronic decay.  The  values of the parameters that  describe  charmonium  are:
\begin{equation}
  k_c = 1.2  \, {\rm GeV } ; \,\,   \sqrt{\Gamma_c } = 0.55  \, {\rm GeV } ; \,\, M_c=2.2  \, {\rm GeV }\,.
  \label{parameters1}
\end{equation} 
Figure \ref{fig:1Tan} presents the effective potentials for the charmonium model at various temperatures. Notice that  the barrier persists at high temperatures.  The collision between the minimum and maximum of the potential occurs only at $T_v \approx 1.4$ GeV in this model. Consequently, the barrier only disappears at $T_v \approx 1.4$ GeV.

\begin{table}[htbp]
  \centering
  \begin{tabular}{|c|c|c||c|c||c|c|}
    \hline
    \multicolumn{1}{|c|}{$n=1$} 
    & \multicolumn{2}{|c||}{Shooting Method} 
    & \multicolumn{2}{|c||}{1st Order -- BS Rule} 
    & \multicolumn{2}{|c||}{Deviation (\%)}  \\
    \cline{2-7}
    $T$ (MeV) & $\omega_R$ (GeV) & $\omega_I$ (GeV)  & $\omega_R$ (GeV)  & $\omega_I$ (GeV)  & $\Delta_R$ & $\Delta_I$  \\
    \hline
    $150$ & $4.55982$ & $1.63822$ & $4.52493$ & $1.63218$ & $7.65 \times 10^{-1}$ & $3.69 \times 10^{-1}$ \\
    $200$ & $5.02645$ & $2.69910$ & $4.99348$ & $2.70329$ & $6.55 \times 10^{-1}$ & $1.55 \times 10^{-1}$ \\
    $250$ & $5.57900$ & $3.78341$ & $5.57765$ & $3.79235$ & $2.42 \times 10^{-2}$ & $2.36 \times 10^{-1}$ \\
    $300$ & $6.17378$ & $4.87560$ & $6.23108$ & $4.90930$ & $9.27 \times 10^{-1}$ & $6.90 \times 10^{-1}$ \\
    $350$ & $6.83430$ & $5.97959$ & $6.93319$ & $6.06018$ & $1.45$ & $1.35$ \\
    $400$ & $7.57351$ & $7.13602$ & $7.68290$ & $7.24973$ & $1.45$ & $1.59$ \\
    \hline
  \end{tabular}
  \caption{The frequencies for the ground state, calculated using both the shooting method and the BS formula  (\ref{BSGACCorrections}), are presented for various temperature.}
  \label{tablegroundstatetop1TMTangent}
\end{table}

In Table~\ref{tablegroundstatetop1TMTangent}, we present the quasinormal frequencies for the second mode, calculated using the WKB approximation~(\ref{BSGACCorrections}). For comparison, we also provide the values obtained using the shooting method. Observe that the two methods show good agreement.

We also present the quasinormal  frequencies for the third mode in Table \ref{tablegroundstatetop1TMTangent2}. Similarly, the results obtained using the WKB approximation show good agreement with those found via the shooting method, as indicated by the deviation values. This demonstrates that the WKB approximation provides accurate results for larger width values, as observed in the case of the second mode.

\begin{table}[htbp]
  \centering
  \begin{tabular}{|c|c|c||c|c||c|c|}
    \hline
    \multicolumn{1}{|c|}{$n=2$} 
    & \multicolumn{2}{|c||}{Shooting Method} 
    & \multicolumn{2}{|c||}{1st Order -- BS Rule} 
    & \multicolumn{2}{|c||}{Deviation$\%$}\\
    \cline{2-7}
    $T$ (MeV) & $\omega_R$ (GeV)  & $\omega_I$ (GeV)  & $\omega_R$ (GeV)  & $\omega_I$ (GeV)  & $\Delta_R$  & $\Delta_I$ \\
    \hline
    $100$ & $4.95792$ & $1.18387$ & $4.93877$ & $1.17689$ & $3.86 \times 10^{-1}$ & $5.89 \times 10^{-1} $ \\
    $120$ & $5.13390$ & $1.79203$ & $5.11284$ & $1.78878$ & $4.10\times 10^{-1}$ & $1.82 \times 10^{-1}$ \\
    $140$ & $5.38116$ & $2.39164$ & $5.36159$ & $2.39318$ & $3.68 \times 10^{-1} $ & $6.43\times 10^{-2}$ \\
    $160$ & $5.67452$ & $2.98502$ & $5.65919$ & $2.99254$ & $2.70\times 10^{-1}$ & $2.52 \times 10^{-1}$ \\
    $180$ & $5.99731$ & $3.57239$ & $5.99120$ & $3.58994$ & $1.02 \times 10^{-1}$ & $4.91 \times 10^{-1}$ \\
    $200$ & $6.30485$ & $4.18534$ & $6.34875$ & $4.18827$ & $6.95 \times 10^{-1}$ & $7 \times 10^{-2}$ \\
    \hline
  \end{tabular}
  \caption{The frequencies for the ground state, calculated using both the shooting method and the BS formula  (\ref{BSGACCorrections}), are presented for various temperature.}
  \label{tablegroundstatetop1TMTangent2}
\end{table}

\section{Discussion and Conclusion}

In this work, we applied the WKB approximation to derive a generalized Bohr-Sommerfeld  quantization rule in  holographic QCD models. Using this formalism, we computed the quasinormal modes  for scalar and vector fields in soft-wall models. For comparison, we also computed the quasinormal modes using the shooting method. The results derived via the BS quantization rule are in good agreement with those obtained using the shooting method, including for the ground state.


The generalized BS quantization rules employed in this work can be applied to other holographic models, provided that the corresponding effective potential exhibits both a well and a barrier. However, in certain holographic QCD setups, it is not possible to find modes localized near the potential minimum due to the existence of a minimum temperature. An explicit example is given by the model discussed in Ref.~\cite{Mamani:2022qnf}, where a black hole solution exists only for temperatures above a certain critical value. As a result, for temperatures above this minimum threshold, the quasinormal modes are localized either near the top of the potential barrier or in the above-barrier region.


Unlike other numerical techniques commonly used in holographic QCD models, such as the power series method~\cite{Horowitz:1999jd,Miranda:2009uw,Mamani:2013ssa}, the Breit-Wigner method~\cite{Berti:2009wx}, and the pseudospectral method~\cite{boyd2001chebyshev,Jansen:2017oag}, the WKB approximation offers valuable information on the dissociation process. Specifically, the shift of quasinormal modes into the above-barrier region can be interpreted as a signal of particle dissociation in the dual field theory, consistent with the behavior observed in the spectral function. However, in the high-temperature regime, the effective potential develops unusual features: At a certain temperature, the barrier disappears entirely. Despite this, the classical turning points remain unchanged, suggesting that the BS formula~(\ref{BSGACCorrections}) may still be applicable. This highlights a limitation of the WKB approximation, as it does not fully capture the dynamics associated with the vanishing of the barrier. Note that other methods commonly used to compute quasinormal modes in holography also fail to describe this transition. Moreover, the quasinormal modes in this regime resemble those of conformal $AdS_5$ spacetime, with frequencies exhibiting linear scaling with temperature. This similarity suggests that the above-barrier region effectively corresponds to the conformal limit.

Finally, we remark that the BS quantization rules given by Eqs.~(\ref{BSqss}) and~(\ref{BSG2}) have already been employed to compute the quasinormal modes of black holes; see, for instance, Refs.~\cite{Guo:2021enm,Volkel:2017ofl}. However, these applications were limited to potentials without an infinite barrier. In the context of gauge/gravity duality, an application of a BS-type quantization formula to compute quasinormal modes can be found in Ref.~\cite{Festuccia:2008zx}. Further examples of WKB approximations applied to systems without a finite barrier are provided in Refs.~\cite{Schutz:1985km, mashhoon_2025_wcbwj-29z57, Konoplya:2003ii, Schutz:1985zz, Iyer:1986np, Iyer:1986nq, Kokkotas:1987pd, Seidel:1989bp, Guinn:1990ic, Mashhoon:1983fg, Mashhoon:1985cya, Mashhoon:1982zz, Mashhoon:1986, Mashhoon:1989, Mashhoon:1992, Mashhoon:1996, Mashhoon:1999, Mashhoon:2002}.

\paragraph*{\textbf{Acknowledgments}: We would like to acknowledge Diego M. Rodrigues for discussions along the development of this work.  L.F.F is supported by ANID Fondecyt postdoctoral grant folio No. 3220304.}

\appendix

\section{Shooting Method}\label{SM}

The shooting method~\cite{Kaminski:2009ce} consists of specifying two boundary conditions at the horizon and then adjusting the free parameter, which is the frequency. In the case of the scalar field, we need to solve the following equation:
\begin{equation}\label{eqz1A}
    \phi''+\phi'\left( \frac{f'}{f} - \frac{3}{z}-\Phi'\right)+\left(\frac{\omega^{2}}{f^2}-\frac{q^2}{f}\right) \phi = 0.
\end{equation}
 using the boundary conditions given by the infalling behavior at the horizon:
 \begin{eqnarray}
     \phi(z\rightarrow z_h)=\left(1-\frac{z}{z_h}\right)^{-i\omega/4\pi T}\left(1+a_1\left(1-\frac{z}{z_h} \right) +a_2\left(1-\frac{z}{z_h} \right)^2+\cdots\right)
 \end{eqnarray}
where the parameters $a_1$ and $a_2$ are given by:
\begin{eqnarray}
   && a_1=\frac{z_h \left(4 i q^2 z_h-3 i \omega^2 z_h-4 \omega z_h \phi '(z_h)-6 \omega\right)}{8 (2 i+\omega z_h)} \\ \cr
 && a_2 = \frac{z_h}{128 (\omega  z_h+2 i) (\omega z_h+4 i)}\bigg(4z_h\omega(2i+z_h\omega)\phi''(z_h)-49z_h\omega^2-2iz_h^2\omega^3-z_h^3(4q^2-3\omega^2)^2\nonumber \\  \cr &&-32i\omega+8z_h\left(\phi'(z_h)\left(8q^2z_h+i\omega(24+z_h^2(-4q^2+3\omega^2)\right)+2z_h\omega(4i+z_h\omega)\phi'(z_h)\right)\bigg)
\end{eqnarray}
The remaining coefficients in the infalling expansion can be obtained by inserting this ansatz into the equation of motion and requiring regularity at the horizon. In this expansion, we consider terms up to $\mathcal{O}\left(1 - z/z_h\right)^{12}$.
Thus, by imposing boundary conditions at the horizon, one can determine a quasinormal mode by varying the frequency until the field satisfies the Dirichlet condition at the boundary.

The same procedure can be applied to the vector case, whose equation of motion (in the absence of momentum) is given by: \begin{equation}\label{eqVector} A_x'' + A_x'\left( \frac{f'}{f} - \frac{3}{z} - \Phi' \right) + \left( \frac{\omega^{2}}{f^2} - \frac{q^2}{f} \right)A_x = 0, \end{equation} where the vector field must satisfy the infalling condition at the horizon. The expansion coefficients are obtained by substituting the ansatz into the equation of motion and imposing regularity at the horizon.

\section{Boundary conditions of WKB wave solutions}\label{BSHB}

To ensure that the WKB wave function satisfies the boundary conditions correctly both near the boundary  and near the horizon  it is essential to analyzes  the WKB wave function in the region $r_*>r_{0*}$ and $r_h<r_*<r_{2*}$ respectively. The computation can be handled analytically by expanding in temperature, particularly in the low-temperature regime, where the integrals are solvable using asymptotic methods.

Thus, to compute the integrals analytically, we return to the holographic coordinate using the relation~(\ref{TL}). We then analyze the following expression near the boundary: \begin{equation} \Psi(z) = \frac{\sqrt{i}, C}{2\sqrt{p_L(z)}} \exp\left[i \int p_L(z') \frac{dz'}{f(z')} \right], \qquad z < z_{0}. \end{equation} This integral can be solved perturbatively in terms of the temperature.
Expanding the integral up to fourth order in temperature and evaluating the expression near the boundary yields: \begin{eqnarray} \Psi(z \rightarrow 0) \sim z^{5/2} \left( d_0(\varepsilon_n, T) + z^{2} d_1(\varepsilon_n, T) + \cdots \right), \end{eqnarray} 
where $d_0$ and $d_1$ are coefficients that depend on both the temperature and the energy parameter $\varepsilon_n$. Note that this solution exhibits the same asymptotic behavior as the normalizable solution in Eq.~(\ref{bb}), while the non-normalizable contribution is absent from the expansion.

Now, considering the WKB wave function near the horizon: \begin{eqnarray} \Psi(z) = \frac{C_1}{\sqrt{p_L(z)}} \exp\left[i \int_{z_z}^{z} p_L(z') \frac{dz'}{f(z')} \right], \qquad z_{2} < z < z_{h}, \end{eqnarray} and expanding this expression near the horizon, we obtain: \begin{eqnarray} \Psi(z \rightarrow z_h) \sim e^{-i \omega r} \left( c_0 + \cdots \right), \end{eqnarray} where $c_0$ is a coefficient that depends on both the frequency and the temperature. 

Therefore, the WKB wave function satisfies the required boundary conditions at both the horizon and the boundary. This behavior is consistent with the boundary conditions for quasinormal modes in AdS space, where the solution must represent an ingoing wave at the horizon and satisfy a Dirichlet condition near the boundary.

\section{Spectral Function}\label{SpectralFunction}

We now derive the spectral function for a scalar field within the framework of holography, following the holographic prescription \cite{Son:2002sd}. The on-shell  version  of scalar field action is given by:
\begin{equation}\label{acsons}
    S_{on-shell}=\int d^4x \sqrt{-g}e^{-\Phi}g^{zz} \phi \partial_z \phi
\end{equation}
where the boundary term at the horizon has been neglected. Applying a Fourier transform to the scalar field, the action (\ref{acsons})  becomes:
\begin{equation}\label{acsons2}
    S_{on-shell}=\int \frac{d^4k}{(2\pi)^4} \sqrt{-g}e^{-\Phi}g^{zz} \bar{\phi}(z,k) \partial_z \bar{\phi}(z,k)\,.
\end{equation}
To proceed,  we decompose the on-shell scalar field $\bar{\phi}(z,k)$ as:
\begin{equation}\label{decomp}
    \bar{\phi}(z,k)=\phi_k(z)\phi_0(k)
\end{equation}
where $\phi_k(z)$ is the bulk-to-boundary propagator satisfying the condition $\lim_{z\rightarrow0}\phi_k(z)=1$. In addition, the bulk-to-boundary propagator must satisfy an infalling condition at the horizon, as required by the holographic prescription \cite{Son:2002sd}.

Now, substituting the relation (\ref{decomp}) into the on-shell action (\ref{acsons2}), we obtain
\begin{equation}\label{acsons3}
    S_{on-shell}=\int \frac{d^4k}{(2\pi)^4} \phi_0(-k)\mathcal{F}(k,z)\phi_0(k),
\end{equation}
where 
\begin{equation}\label{funcF}
   \mathcal{F}(k,z)=\sqrt{-g}g^{zz} e^{-\Phi}\bar{\phi}^{*}_{k}(z) \partial_z \bar{\phi}_{k}(z)\,.
\end{equation}
The retarded Green's functions is  extracted using the  relation \cite{Son:2002sd}:
\begin{equation}\label{GreenFunction}
G^{R}(k)=-2\mathcal{F}(k,z)
\end{equation}
Finally, the spectral function  is calculated as:
\begin{equation}\label{spectral}
\rho(\omega,q)=-2\, Im \, G^{R}(k)=-4 \, Im \, \mathcal{F}(k,z)
\end{equation}
Hence, by numerically solving the equation of motion with the appropriate boundary conditions at the horizon and the boundary, and applying this relation, the spectral function is obtained.

\section{The Bohr-Sommerfeld quantization at zero temperature }\label{BSzeroT}

\subsection{Scalar Field}

At zero temperature, the holographic potential (\ref{potential1}) for the scalar field takes the  following form:
\begin{equation}
    V(z)=2c^2+\frac{15}{4z^2}+c^4z^2\,.
\end{equation}
\begin{figure}[!htb]
	\centering
	\includegraphics[scale=0.45]{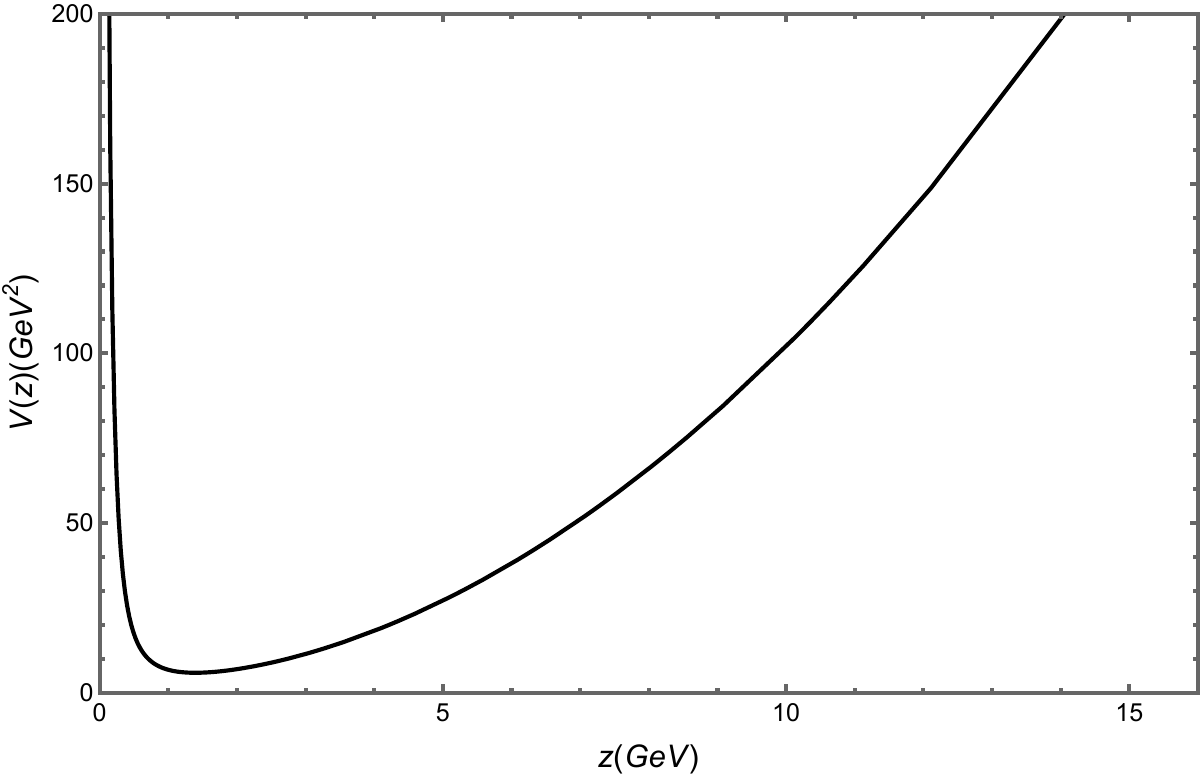}
	\caption{ The  holographic potential for the scalar field at zero temperature as a function of the  holographic coordinate.}
    \label{HolographicPT0}
\end{figure}
We plot in Fig. \ref{HolographicPT0} the holographic potential as a function of the  holographic coordinate. The potential exhibits an infinite barrier near the boundary and does not have a barrier; it is solely a well potential. In this situation, it is possible to apply the BS quantization:
\begin{equation}\label{BSqssT0}
    \int_{z_0}^{z_1}\sqrt{m_n^2-V_L(z)}dz=\pi \left( n+\frac{1}{2}\right)\,.
\end{equation}
where Langer the potential is defined as
\begin{equation}
    V_L(z)=2c^2+\frac{1}{4z^2}+c^4z^2\,,
\end{equation}
with the turning points given by
\begin{eqnarray}
    && z_0=\frac{\sqrt{-2c^2+m_n^2-\sqrt{-12c^4-4m^2_nc^2+m_n^4}}}{\sqrt{2}c^2}\\ \cr
    && z_1=\frac{\sqrt{-2c^2+m_n^2+\sqrt{-12c^4-4m^2_nc^2+m_n^4}}}{\sqrt{2}c^2}\,.
\end{eqnarray}
Thus, by solving the integral in the BS relation, one finds
\begin{equation}
\int_{z_0}^{z_1}\sqrt{m_n^2-V_L(z)}dz=-\frac{3\pi}{2}+\frac{m_n^2\pi}{4c^2},
\end{equation}    
Finally, substituting into the BS expression give us
\begin{equation}
    m_n^2=4c^2(n+2)\,.
\end{equation}
This corresponds exactly to the spectrum of glueball particles in the soft-wall model. It is important to note that applying the Langer transformation is necessary to obtain the correct spectrum. An alternative approach to deriving this spectrum involves incorporating all WKB corrections \cite{MSeetharaman_1984}, or solving the Schrödinger equation near the boundary and then performing a matching with the solutions in other regions, as discussed in \cite{Koike_2009,Karnakov_2013}.

\subsection{Vector Field}

As in the case of scalar field we can use the BS quantization rule:
\begin{equation}\label{BSqssV}
    \int_{z_0}^{z_1}\sqrt{m_n^2-V_L(z)}dz=\pi \left( n+\frac{1}{2}\right)\,.
\end{equation}
Then, by considering the Langer potential in the case of the vector field,
\begin{equation} V_L(z) = \frac{1}{z^2} + c^4 z^2, \end{equation}
with the turning points given by
\begin{eqnarray} && z_0 = \frac{\sqrt{m_n^2 - \sqrt{-4c^4 + m_n^4}}}{\sqrt{2}c^2},\ \cr && z_1 = \frac{\sqrt{m_n^2 + \sqrt{-4c^4 + m_n^4}}}{\sqrt{2}c^2}, \end{eqnarray}
and by solving Eq. (\ref{BSqssV}), one finds
\begin{equation} \int_{z_0}^{z_1} \sqrt{m_n^2 - V_L(z)}, dz = -\frac{\pi}{2} + \frac{m_n^2\pi}{4c^2}. \end{equation}
Hence the spectrum is given by
\begin{equation}
    m_n^2=4c^2(n+1)\,.
\end{equation}

\section{The Bohr-Sommerfeld quantization at very low temperature }\label{BSzerverylowtemperature}
In the low-temperature regime, where the quasinormal modes  are located near the potential minimum, an analytical expression for the real part of the frequencies can be derived in the case of the original soft-wall model \cite{Karch:2006pv}. To illustrate this, let us first examine the case of a scalar field. 

The equation (\ref{BScorrections}) can be expressed in terms of the  $z$ coordinate using the transformation given by (\ref{TL}). Consequently, the BS formula (\ref{BScorrections}) can be rewritten as follows:
\begin{equation}\label{BScorrectionsz}
     \int_{z_0}^{z_1}\sqrt{\varepsilon_R-V_L(z)}\frac{dz}{f}=\pi \left( n+\frac{1}{2}\right)+I_1
\end{equation}
where

\begin{eqnarray}
I_1&=& -\frac{1}{12} \frac{d^2}{d\varepsilon^2_R}\int_{z_0}^{z_1} \Bigg[\bigg( -24\pi T\left(\varepsilon_{R}-V_{L}\right)\left[4\pi T (V_L-V_T)-V_T'f(\arctan(\pi Tz)+ \operatorname{arctanh}(\pi T z)\right] \nonumber \\ \nonumber \cr  &+&16\pi^2T^2\left(\varepsilon_{R}-V_{L}\right)^2 \left(-4\pi T\left(V_L-V_T\right)+V_T'f\left(\arctan(\pi Tz)+ \operatorname{arctanh}(\pi T z)\right)\right)^2 \bigg) \\ \cr  && \times\frac{1}{\left(\arctan(\pi Tz)+ \operatorname{arctanh}(\pi T z)\right)^2f\sqrt{\varepsilon_{R}-V_L}} \Bigg]
\end{eqnarray}
with
\begin{align}
\label{potential2_VT}
V_T(z) &= 2c^2 + c^4z^2 
- \pi^4 T^4 \left( \frac{3}{2}z^2 + 2c^4z^6 \right) 
+ \pi^8 T^8 \left( c^4z^{10} - \frac{9}{4}z^6 - 2c^2z^8 \right), \\\nonumber \\
\label{potential2_VL}
V_L(z) &= 2c^2 + \frac{15}{4z^2} + c^4z^2 
- \pi^4 T^4 \left( \frac{3}{2}z^2 + 2c^4z^6 \right) 
+ \pi^8 T^8 \left( c^4z^{10} - \frac{9}{4}z^6 - 2c^2z^8 \right) \notag \\
&\quad + \frac{\pi^2 T^2}{\left( \operatorname{arctan}(\pi T z) + \operatorname{arctanh}(\pi T z) \right)^2}.
\end{align}

The turning points are determined using the relation $\varepsilon_R-V_L=0$. The results are shown below up to the fourth order:
\begin{eqnarray}
z_0=&&\frac{\sqrt{\varepsilon_R - \sqrt{(\varepsilon_R - 6c^2)(\varepsilon_R + 2c^2)} - 2c^2}}{\sqrt{2}}+\frac{\pi^4 T^4 \left(4 \varepsilon_R^3 - 24c^2 \varepsilon_R^2 + 19 c^4\varepsilon_R + 26c^6\right) }{4 c^{10}\sqrt{2} \sqrt{\varepsilon_R - \sqrt{(\varepsilon_R - 6c^2)(\varepsilon_R + 2c^2)} - 2c^2}} + \nonumber \\ \cr &&+ \frac{\pi ^4 T^4} { \sqrt{(\varepsilon_R - 6c^2)(\varepsilon_R + 2c^2)}} \times\frac{ \left(-4 \varepsilon_R^4 + 32c^2 \varepsilon_R^3- 35 c^4\varepsilon_R^2 - 116 c^6\varepsilon_R + 76c^{8}\right) }{4c^{10} \sqrt{2} \sqrt{\varepsilon_R - \sqrt{(\varepsilon_R - 6c^2)(\varepsilon_R + 2c^2)} - 2c^2}}
\end{eqnarray}

\begin{eqnarray} z_1=&&\frac{\sqrt{\varepsilon_R + \sqrt{(\varepsilon_R - 6c^2)(\varepsilon_R - 2c^2)} - 2c^2}}{\sqrt{2}c^2}+\frac{\pi ^4 T^4 \left(4 \varepsilon_R^3 - 24c^2 \varepsilon_R^2 + 19c^4 \varepsilon_R + 26c^6\right)}{4c^{10} \sqrt{2} \sqrt{\varepsilon_R + \sqrt{(\varepsilon_R - 6c^2)(\varepsilon_R + 2c^2)} - 2c^2}} +\nonumber \\ \cr && -\frac{\pi ^4 T^4} { \sqrt{(\varepsilon_R - 6c^2)(\varepsilon_R + 2c^2)}}\times \frac{ \left(-4 \varepsilon_R^4 + 32 c^2 \varepsilon_R^3- 35 c^4\varepsilon_R^2 - 116c^6 \varepsilon_R + 76c^8\right) }{4 \sqrt{2} c^{10}\sqrt{\varepsilon_R + \sqrt{(\varepsilon_R - 6c^2)(\varepsilon_R + 2c^2)} - 2c^2}}\end{eqnarray}

\begin{eqnarray} && z_2 =\frac{1}{\pi T}-\frac{\sqrt{\varepsilon_R}}{4c^2} + \left(\frac{1}{2c^2} - \frac{5 \varepsilon_R}{32c^4}\right) \pi T + \left(-\frac{15}{32c^2 \sqrt{\varepsilon_R}} + \frac{\sqrt{\varepsilon_R}}{2c^4} - \frac{5}{32c^6} \varepsilon_R^{3/2} \right) (\pi T)^2 \\ \cr && + \left(\frac{7}{128c^4} + \frac{49 \varepsilon_R}{64c^6} - \frac{385 \varepsilon_R^2}{2048^6}\right) (\pi T)^3 + \left(\frac{225}{512 c^2\varepsilon_R^{3/2}} - \frac{35}{32c^4\sqrt{\varepsilon_R}} - \frac{9 \sqrt{\varepsilon_R}}{16c^6} + \frac{5}{4c^8} \varepsilon_R^{3/2} - \frac{1}{4c^{10}} \varepsilon_R^{5/2} \right) (\pi T)^4 \nonumber \end{eqnarray}
Now, by substituting the turning points into formula (\ref{BScorrectionsz}), the integrals are evaluated. Additionally, we approximate the real part of the frequency as $\omega_R \approx \sqrt{\varepsilon_R}$, since, in this regime, the imaginary part is very small. Thus, we obtain the following expression for the real part of the frequency:
\begin{eqnarray} \label{massTScalar}
\omega_R \approx \sqrt{\varepsilon_R} \approx &&  \frac{\sqrt{n+2}}{4}\Bigg(8c-\frac{24}{c^3}  (n+1) (2 n+3) \pi ^4T^4 -\frac{\pi ^8T^8 }{90c^7(n+2)} \bigg( 134640 n^5+1015920 n^4+3076920 n^3 \cr &&  +4677840 n^2+3543707 n+1052733\bigg)\Bigg)\dots +O(T^{12})\,.
 \end{eqnarray}
Here, we only considered until the 8th order, but this  is enough to show that the results are in agreement with numerical results,  as can be observed in Table \ref {TableAnalS}. 

\begin{table}[htbp]
  \centering
  \begin{tabular}{||c||c|c||c|c||c||}
    \hline
    \multicolumn{1}{||c||}{n=0} &\multicolumn{1}{|c||}{Shooting Method} &\multicolumn{1}{|c||}{Numerical BS rule }&\multicolumn{1}{|c||}{Analytic BS rule } \\
     \cline{1-4}
     \multicolumn{1}{||c||}{T(MeV)}  & $\omega_R$ (GeV)   & $\omega_R$ (GeV)   & $\omega_R$ (GeV)    \\\hline
  $20$  &  $2.82803 $   &  $2.82803$   &  $2.82803$    \\  \cline{2-4}    
    $25$  & $2.82745$ &    $2.82745$  &   $2.82745$   \\ \cline{2-4}
    $30$  & $2.82640$    &  $2.82640$    &    $2.82640$  \\ \cline{2-4}
    $35$  & $2.82194$ &    $2.82194$    & $2.82466$  \\ \cline{2-4}\hline
 \end{tabular}
  \caption{The real part of the frequencies for the ground state of the glueball particle with 
$c=1$, computed using both the shooting method and the WKB approximation evaluated both numerically and analytically are presented for a range of temperatures.}
  \label{TableAnalS}
\end{table}

In case of the vector field, the holographic potential is given 
\begin{align}
\label{potential2_VTVector}
V_T(z) &=  c^4z^2 
+ \pi^4 T^4 \left( \frac{1}{2}z^2 +4c^2z^4- 2c^4z^6 \right) 
+ \pi^8 T^8 \left( c^4z^{10} - \frac{5}{4}z^6 - 4c^2z^8 \right), \\\nonumber \\
\label{potential2_VL}
V_L(z) &=  \frac{3}{4z^2} + c^4z^2 
- \pi^4 T^4 \left( \frac{1}{2}z^2 +4c^2z^4- 2c^4z^6 \right) 
+ \pi^8 T^8 \left( c^4z^{10} - \frac{5}{4}z^6 - 4c^2z^8 \right) \notag \\
&\quad + \frac{\pi^2 T^2}{\left( \operatorname{arctan}(\pi T z) + \operatorname{arctanh}(\pi T z) \right)^2}.
\end{align}
and  the turning points are given by 
\begin{eqnarray}\label{TPVz0}
z_0=&&\frac{\sqrt{\varepsilon_R- \sqrt{\varepsilon_R^2 - 4c^4}}}{\sqrt{2}c^2}-\frac{\pi^4 T^4 \left(4 \varepsilon_R^4-8c^2 \varepsilon_R^3 - 17c^4 \varepsilon_R^2 + 24 c^6\varepsilon_R + 10c^8\right) }{4c^{10} \sqrt{2} \sqrt{\varepsilon_R^2-4c^4} \sqrt{\varepsilon_R - \sqrt{\varepsilon_R^2 - 4c^4}}} + \frac{\pi ^4 T^4}{\varepsilon_R^2-4c^4}  \nonumber \\ \cr \times&& \frac{ \left(4 \varepsilon_R^5 -8c^2\varepsilon_R^4- 25 c^4 \varepsilon_R^3+40 c^6\varepsilon_R^2 + 36 c^8\varepsilon_R -32c^{10}\right) }{4c^{10}\sqrt{2}  \sqrt{\varepsilon_R - \sqrt{\varepsilon_R^2 - 4c^4}}}
\end{eqnarray}

\begin{eqnarray}\label{TPVz1}
z_1=&&\frac{\sqrt{\varepsilon_R + \sqrt{\varepsilon_R^2 - 4c^4}}}{\sqrt{2}c^2}+\frac{\pi^4 T^4 \left(4 \varepsilon_R^4-8c^2 \varepsilon_R^3 - 17 c^4\varepsilon_R^2 + 24 c^6\varepsilon_R + 10c^8\right) }{4 c^{10} \sqrt{2} \sqrt{\varepsilon_R^2-4c^4} \sqrt{\varepsilon_R + \sqrt{\varepsilon_R^2 - 4c^4}}} + \frac{\pi ^4 T^4}{\varepsilon_R^2-4c^4}  \nonumber \\ \cr \times&& \frac{ \left(4 \varepsilon_R^5 -8c^2\varepsilon_R^4- 25c^4\varepsilon_R^3+40 c^6\varepsilon_R^2 + 36c^8\varepsilon_R -32c^{10}\right) }{4 c^{10}\sqrt{2}  \sqrt{\varepsilon_R + \sqrt{\varepsilon_R^2 - 4c^4}}}
\end{eqnarray}

\begin{eqnarray}\label{TPVz2} && z_2 =\frac{1}{\pi T}-\frac{\sqrt{\varepsilon_R}}{4c^2} + \left(\frac{1}{2c^2} - \frac{5 \varepsilon_R}{32c^4}\right) \pi T + \left(-\frac{15}{32 c^2 \sqrt{\varepsilon_R}} + \frac{\sqrt{\varepsilon_R}}{4c^4} - \frac{5}{32c^6} \varepsilon_R^{3/2} \right) (\pi T)^2+ \\ \cr && + \left(\frac{71}{128c^4} + \frac{21 \varepsilon_R}{64c^6} - \frac{385 \varepsilon_R^2}{2048c^8}\right) (\pi T)^3 + \left(\frac{225}{ 512c^2\varepsilon_R^{3/2}} - \frac{25}{16c^4 \sqrt{\varepsilon_R}} + \frac{7}{16c^6}\sqrt{\varepsilon_R} + \frac{1}{2c^8} \varepsilon_R^{3/2} - \frac{1}{4c^{10}} \varepsilon_R^{5/2} \right) (\pi T)^4. \nonumber \end{eqnarray}

As before, we need only to insert the holographic potential (\ref{potential2_VTVector}) and the turning points (\ref{TPVz0}) and (\ref{TPVz1}) into expression (\ref{BScorrectionsz}) and solve it. This yields the real part of the frequency for the vector field:
\begin{eqnarray} \label{massTVector}
\omega_R \approx \sqrt{\varepsilon_R} \approx &&  \frac{\sqrt{n+1}}{20}\Bigg(40c-\frac{3}{c^3}\pi^4 T^4  (53+120n+80n^2)  -\frac{\pi^8 T^8 }{720c^7(n+1)} \bigg( 5385600 n^5+22521600 n^4 \cr && +41140400 n^3 +40430400 n^2+20155811 n+3841771\bigg)\Bigg)\dots +O(T^{12})\,.
 \end{eqnarray}
\begin{table}[htbp]
  \centering
  \begin{tabular}{||c||c|c||c|c||c||}
    \hline
    \multicolumn{1}{||c||}{n=0} &\multicolumn{1}{|c||}{Shooting Method} &\multicolumn{1}{|c||}{Numerical BS rule }&\multicolumn{1}{|c||}{Analytic BS rule } \\
     \cline{1-4}
     \multicolumn{1}{||c||}{T(MeV)}  & $\omega_R$ (GeV)   & $\omega_R$ (GeV)   & $\omega_R$ (GeV)    \\\hline
  $20$  &  $1.99988 $   &  $1.99988 $   &  $1.99988$   \\  \cline{2-4}    
    $25$  & $1.99970$ &    $1.99970$  &   $1.99970$   \\ \cline{2-4}
    $30$  & $1.99937$    &  $1.99937$    &    $1.99937$  \\ \cline{2-4}
    $35$  & $1.99882$ &    $1.99882$    & $1.99883$ \\ \cline{2-4}\hline
 \end{tabular}
  \caption{The real part of the frequencies for the ground state of the vector meson particle with 
$c=1$, computed using both the shooting method and the WKB approximation evaluated both numerically and analytically are presented for a range of temperatures.}
  \label{TableAnalV}
\end{table}
For comparison, we present in the table \ref{TableAnalV} the results obtained using the previously discussed numerical procedure. As evident, the formulas (\ref{massTVector}) and (\ref{massTScalar}) provide an excellent approximation of the real part of the frequency at very low temperatures. Moreover, the accuracy can be further improved by including additional terms in the series expansion.

\bibliographystyle{apsrev4-2}
\bibliography{refs}

\end{document}